\documentclass[showpacs,twocolumn,aps,superscriptaddress]{revtex4}
\usepackage{dcolumn}
\usepackage{bm}
\usepackage{graphicx, subfigure}

\DeclareGraphicsExtensions{.pdf,.png,.jpg}

\begin{document}

\newcommand{\vAi}{{\cal A}_{i_1\cdots i_n}}
\newcommand{\vAim}{{\cal A}_{i_1\cdots i_{n-1}}}
\newcommand{\vAbi}{\bar{\cal A}^{i_1\cdots i_n}}
\newcommand{\vAbim}{\bar{\cal A}^{i_1\cdots i_{n-1}}}
\newcommand{\br}{bremsstrahlung }
\newcommand{\Br}{Bremsstrahlung}
\newcommand{\htS}{\hat{S}}
\newcommand{\htR}{\hat{R}}
\newcommand{\htB}{\hat{B}}
\newcommand{\htD}{\hat{D}}
\newcommand{\htV}{\hat{V}}
\newcommand{\cT}{{\cal T}}
\newcommand{\cM}{{\cal M}}
\newcommand{\cMs}{{\cal M}^*}
\newcommand{\vk}{{\bf k}}
\newcommand{\vK}{{\bf K}}
\newcommand{\vb}{{\bf b}}
\newcommand{{\vp}}{{\bf p}}
\newcommand{{\vq}}{{\bf q}}
\newcommand{\vQ}{{\bf Q}}
\newcommand{\vx}{{\bf x}}
\newcommand{\tr}{{{\rm Tr}}}
\newcommand{\beq}{\begin{equation}}
\newcommand{\eeq}[1]{\label{#1} \end{equation}}
\newcommand{\half}{{\textstyle \frac{1}{2}}}
\newcommand{\gton}{\stackrel{>}{\sim}}
\newcommand{\lton}{\mathrel{\lower.9ex
                  \hbox{$\stackrel{\displaystyle <}{\sim}$}}}
\newcommand{\ee}{\end{equation}} \newcommand{\ben}{\begin{enumerate}}
\newcommand{\een}{\end{enumerate}} \newcommand{\bit}{\begin{itemize}}
\newcommand{\eit}{\end{itemize}} \newcommand{\bc}{\begin{center}}
\newcommand{\ec}{\end{center}} \newcommand{\bea}{\begin{eqnarray}}
\newcommand{\eea}{\end{eqnarray}}
\newcommand{\beqar}{\begin{eqnarray}}
\newcommand{\eeqar}[1]{\label{#1} \end{eqnarray}}
\newcommand{\bra}[1]{\langle {#1}|}
\newcommand{\ket}[1]{|{#1}\rangle}
\newcommand{\norm}[2]{\langle{#1}|{#2}\rangle}
\newcommand{\brac}[3]{\langle{#1}|{#2}|{#3}\rangle}
\newcommand{\hilb}{{\cal H}}
\newcommand{\pleft}{\stackrel{\leftarrow}{\partial}}
\newcommand{\pright}{\stackrel{\rightarrow}{\partial}}

\title{ Non-Abelian Bremsstrahlung and Azimuthal Asymmetries \\ 
in High Energy p+A  Reactions}

\author{M. Gyulassy}
\email[*]{gyulassy@phys.columbia.edu}
\affiliation{MTA WIGNER Research Centre for Physics, RMI, Budapest, Hungary }  
\affiliation{Department of Physics, Columbia University, New York, NY 10027, USA}

\author{P. Levai}
\affiliation{MTA WIGNER Research Centre for Physics, RMI, Budapest, Hungary }  

\author{I. Vitev}
\affiliation{Theoretical Division, Los Alamos National Laboratory, Los
Alamos, NM 87545, USA }  
\author{T. Biro}

\affiliation{MTA WIGNER Research Centre for Physics, RMI, Budapest, Hungary }

\date{May 30, 2014}
                                                                              
\begin{abstract}
We apply the GLV reaction operator solution to the
Vitev-Gunion-Bertsch (VGB) boundary conditions to compute the
all-order in nuclear opacity non-abelian gluon \br of event-by-event
fluctuating beam
jets in nuclear collisions. We evaluate analytically azimuthal
Fourier moments  of single gluon, $v_n^M\{1\}$, 
and even number $2\ell$  gluon, $v_n^M\{2\ell\}$
inclusive distributions in high
energy p+A reactions as a function of harmonic $n$, 
target recoil cluster number, $M$, and gluon number, $2\ell$,
at  RHIC and LHC.  Multiple resolved clusters of 
recoiling target beam jets together with the projectile beam jet form 
Color Scintillation Antenna (CSA) arrays that lead to characteristic
boost non-invariant trapezoidal rapidity distributions in asymmetric
$B+A$ nuclear collisions.  The scaling of intrinsically azimuthally anisotropic 
and long range in $\eta$ nature of the non-abelian \br 
leads to  $v_n$ moments that are similar to results from hydrodynamic models,
but due entirely to non-abelian wave interference phenomena
sourced by the fluctuating CSA. Our analytic non-flow solutions are similar to
recent numerical saturation model predictions but differ by predicting
a simple power-law hierarchy of both even and odd $v_n$ without
invoking $k_T$ factorization.  A test of CSA mechanism is the
predicted nearly linear  $\eta$ rapidity dependence of the $v_n(k_T,\eta)$.
Non-abelian beam jet \br may thus 
provide a simple analytic solution to Beam Energy
Scan (BES) puzzle of the near $\sqrt{s}$ independence of $v_n(p_T)$ 
moments observed down to 10 AGeV where large-$x$ valence quark beam jets
dominate inelastic dynamics. Recoil \br from multiple independent
CSA clusters could also provide a partial explanation
for the unexpected  similarity of $v_n$ in $p(D)+A$ and
non-central $A+A$ at same $dN/d\eta$ multiplicity as observed at RHIC 
and LHC.
\end{abstract}


\maketitle

\section{Introduction}

\begin{figure}[!tb]
\centerline{\includegraphics[width=3.in]{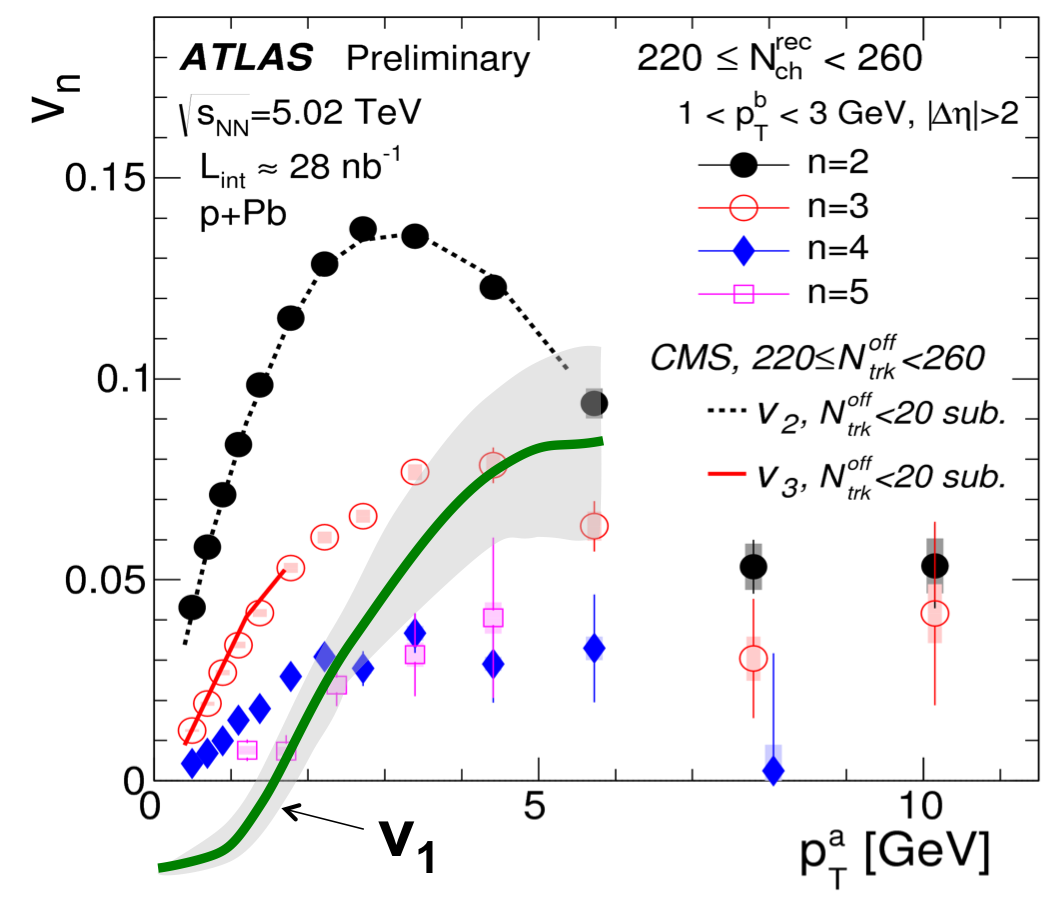}}
\caption{(Color online) Reproduced from ATLAS~\cite{ATLASvn} p+Pb Figure 24
$v_n(p_T)$ with n=2 to 5 obtained for $|\Delta \eta|>2$ and the $p_T$ range of 1-3~GeV. An overlay sketch of preliminary rapidity-even $v_1$ data shown at QM14
\cite{ATLASpAv1} is also indicated.
The error bars and shaded boxes represent 
the statistical and systematic uncertainties, respectively. Results in $220≤N_{ch}<260$
 are compared to the CMS data~\cite{CMS:2012qk} obtained by subtracting the peripheral events 
 (the number of offline tracks $N^{trk}_{off}<20$), shown by the solid and dashed lines. }
\label{ATLASfg24}
\end{figure}

An unexpected discovery at RHIC/BNL in $D+Au$ reactions at $\sqrt{s}=200$ AGeV~\cite{Adare:2013piz}   and 
at LHC/CERN  in $\sqrt{s}=5.02$ ATeV $p+Pb$  reactions~\cite{CMS:2012qk,Abelev:2012ola,Aad:2012gla}
is the large magnitude of mid-rapidity 
azimuthal anisotropy moments, $v_n(k_T, \eta=0)$, that are remarkably 
similar to those observed previously in non-central $Au+Au$~\cite{Adams:2005dq,Adcox:2004mh,Adare:2010ux}
and $Pb+Pb$~\cite{Aamodt:2011by,Aamodt:2010pa,ALICE:2011ab,Chatrchyan:2012wg,ATLAS:2012at}  reactions. See preliminary data in 
Fig.~\ref{ATLASfg24} taken from ATLAS~\cite{ATLASvn} Fig. 24
that also shows a large rapidity-even dipole $v_1$ harmonic\cite{ATLASpAv1}.

In addition, the Beam energy Scan (BES) at RHIC~\cite{Adamczyk:2013gw}
revealed a near $\sqrt{s}$ independence from 8~AGeV to 2.76~ATeV of
the $v_n$ in $A+A$ at fixed centrality that was also unexpected.

In high energy $A+A$, the $v_n$ moments have been interpreted as
possible evidence for the near ``perfect fluidity'' of the
strongly-coupled Quark Gluon Plasmas (sQGP) produced in such
reactions~\cite{Romatschke:2007mq, Luzum:2008cw, Alver:2010dn, Gale:2012rq, Heinz:2013th}.  However, the recent observation of similar $v_n$ in
much smaller $p(D)+A$ systems and also the near beam energy
independence of the $A+A$ moments observed in the Beam Energy Scan (BES)~\cite{Adamczyk:2013gw} at RHIC together with LHC,
from 7.7~AGeV to 2.76~ATeV have posed a problem for the
perfect fluid interpretation because near inviscid hydrodynamics is
not expected to apply in space-time regions where the local
temperature falls below the confinement temperature, $T(x,t) <T_c\sim
160 $ MeV. In that Hadron Resonance Gas (HRG) ``corona'' region, the
viscosity to entropy ratio is predicted to grow rapidly with
decreasing temperature~\cite{Danielewicz:1984ww} and the corona volume
fraction must increase relative to the ever shrinking volume of the
perfect fluid ``core'' with $T>T_c$ when either  the projectile atomic
number $A$ and size $A^{1/3}$ fm or the center-of-mass (CM) energy
$\sqrt{s}$ decrease .

While hydrodynamic equations have been shown to be  
sufficient to describe $p(D)+A$ data
with particular assumptions about initial and freeze-out  conditions~\cite{Bozek:2011if}, its necessity as a 
unique interpretation of the data is not guaranteed. 
This point was underlined recently using a specific initial state 
saturation model~\cite{Dusling:2013oia}  that was shown to be able to fit  $p(D)+A$ 
correlation even $v_n$ moments data  without final state interactions.  
That saturation model has also been used~\cite{Gale:2012rq} to specify
initial conditions for perfect fluid hydrodynamics in  $A+A$.  
However, in $p+A$  such initial conditions for hydrodynamics are not as well-controlled
 because the gluon saturation scale scale, $Q_s(x,A=1)<1$ GeV,  is small and its fluctuations 
 in the transverse plane on sub-nucleon scales are not reliably predicted.

The near independence of $v_n$ moments on beam energy observed in BES~\cite{Adamczyk:2013gw}
at RHIC from 7.7 AGeV to 2760 AGeV pose further serious challenges to
the uniqueness of the perfect fluid  interpretations of the data
because of previous predictions~\cite{Teaney:2000cw} for systematic
reduction of the moments due to the increasing HRG corona.
Those predictions appeared to be confirmed by
SPS~$\sqrt{s}=17$~AGeV data~\cite{Agakichiev:2003gg}. The
most recent BES measurements, however, appear to contradict the diluting role
of the HRG corona. The  HRG corona fraction should  dilute
perfect fluid QGP core flow signatures at lower energies 
unless additional dynamical mechanisms possibly associated with increasing
baryon density accidentally conspire to compensate for
growing HRG corona fraction. Such combination of canceling effects
with $\sqrt{s}$ was demonstrated to be possible using
a specific hybrid hydro+URQMD model~\cite{Auvinen:2013sba} or three fluid 
models~\cite{Ivanov:2014zqa}.
While such hybrid models are {\em sufficient} to explain the BES independence
of $v_2$ data in $A+A$,  the {\em necessity} and, hence, uniqueness of such hybrid 
descriptions are  not guaranteed.

The BES~\cite{Adamczyk:2013gw} data also pose a challenge to color glass condensate (CGC)
gluon saturation model \cite{Gale:2012rq} 
used to specify initial conditions for hydrodynamic flow 
predictions in  $A+A$. This is because
 $Q_s^2$ is predicted to decrease with
$\log (s)$, and thus gluon
saturation-dominated high energy gluon fusion models of initial-state 
dynamics should switch
over into valence quark-diquark dominated inelastic dynamics when partons with
fractional energy $x>0.01$ play the dominant role. At RHIC and lower energies
valence quark and diquark QCD string phenomenology
based on the 
LUND~\cite{Andersson:1986gw} model with (diquark-quark) beam jets 
and its $B+A$ nuclear collision
generalization via  HIJING~\cite{Gyulassy:2003mc} 
can smoothly interpolate between AGS and RHIC
energies. Such  multiple beam jet based approach to $B+A$ naturally accounts,
for example, for the striking long range
triangular,  boost non-invariant, form of $(dN_{pA}/d\eta)/(dN_{pp}/d\eta)$
nuclear enhancement of the final hadron
rapidity density in $p(D)+A$ observed at all CM energies up to
LHC~\cite{Debbeqm14}. By including multiple mini and hard jet production
it can account for the $\sqrt{s} $ growth of $dN_{B+A}/d\eta$ though
at top $\sqrt{s}=200$ AGeV RHIC and at LHC energies 
there is strong evidence for the onset for gluon saturation~\cite{Gyulassy:2004zy} that limits
$2\rightarrow 2$ minijet processes to $p_T >Q_s(x,A)\propto 
A^{1/3}/x^{\lambda}$ that  grows with $A$ and $1/x=\sqrt{s}/(p_T e^{\eta})$.
  
The importance of 
multiple beam jets  with rapidity  kinematics controlled by valence quarks and diquarks was first proposed within the 
Brodsky-Gunion-Kuhn (BGK)
model~\cite{Brodsky:1977de} which is reproduced 
also in the HIJING~\cite{Adil:2005qn} model.  The trapezoidal boost non-invariant
dependence of the local density, $dN/d\eta d^2{\bf x}$, predicted in
\cite{Adil:2005qn} as a function
of the transverse coordinate ${\bf x}$ even in symmetric $A+A$,
may also play an important role in
in the triangular long range $\eta$ dependence of 
$v_2(\eta,\sqrt{s}$) as observed in $Au+Au$ by PHOBOS~\cite{Busza:2004mc}.

In this paper we explore the  possibility that a dynamical 
source that could partially account for the above puzzling azimuthal moment systematics 
may be traced to a basic perturbative QCD (pQCD) feature.
The pQCD based model here extends the opacity $\chi=1$  Gunion-Bertsch~\cite{Gunion:1981qs} 
(GB) perturbative QCD \br used to model for $\pi+\pi\rightarrow g+X$ 
to all orders  in opacity, $e^{-\chi}\sum_{n=1}^\infty\chi^n/n! \cdots$, 
Vitev-Gunion-Bertsch (VGA) multiple interaction pQCD \br 
for applications to $B+A$ nuclear collisions. 
We show that VGA \br naturally leads on an event by event basis to a hierarchy of 
non trivial azimuthal asymmetry moments similar to those observed in 
$p+A$ (see fig.1) and peripheral $A+A$ at fixed $dN/d\eta$~\cite{Chatrchyan:2012wg,Aamodt:2010pa,ATLAS:2012at} . 

A particularly important feature of beam jet non-abelian \br is that it
automatically leads to long range rapidity $\eta$ ``ridge'' correlations 
and to  all even and odd azimuthal $v_n\{2\ell\}$ 
harmonics with $n,\ell=1,2,3,\cdots$.
Conventional Lund string beam 
jet models~\cite{Andersson:1986gw}, as encoded e.g. in HIJING, on the other
hand  neglect recoil induced moderate $p_T$
color \br azimuthal asymmetries.  From the pQCD perspective, 
beam jets are simply arrays of parallel
color antennas that radiate due to multiple soft transverse momentum transfers
$|{\bf q}_i|\sim 1$ GeV
between  participant projectile and $i=1,\cdots, N_T(b)$ target nucleons. 
Many event generators  include {\em $\phi$ averaged}
(azimuthally randomized) bremsstrahlung effects via 
 $\sim \alpha_s/k_T^2$  up to the minijet
scale $k_T < Q_s(x,A)$. In HIJING the ARIADNE~\cite{Ariadne} code 
is used in conjunction with the non-perturbative 
Lund string fragmentation code 
 JETSET~\cite{Sjostrand:1993yb} to incorporate
this effect, while 
highly azimuthally asymmetric hard 
pQCD jets with  $k_T>Q_s(x,A)$ are included via the
PYTHIA~\cite{Sjostrand:1993yb}) code.
In~\cite{Andersson:1986gw} it was emphasized that the 
high string tension of color strings reduces greatly the sensitivity of Lund string fragmentation to 
QCD \br  , and that this is an important infrared safety feature of that non-perturbative 
hadronization phenomenology.   

In $p+A$ multiple collisions, however,
the projectile accumulates multiple transverse momentum kicks (the Cronin effect) from scattering with 
cold nuclear  participants~\cite{Gyulassy:2002yv,Ovanesyan:2011xy}
that enhances the \br mean square $<k_T^2>_{pA}\approx A^{1/3} \mu^2 $ via random walk in the 
target frame. In the CGC approach this $A^{1/3}$ growth is built into
 $Q_s^2(x,A)$ in the infinite momentum frame.

At the minijet scale the underlying azimuthal asymmetry
 of non-abelian \br will tend to focus gluons toward
the azimuthal directions of  exchanged momenta. At present,
this basic azimuthal dependence is not taken into account in HIJING. 

As we show below, there is a very important aspect to the  
multiple color antenna arrays in high energy
$p+A$  due to the  longitudinal coherence of clusters of participant target beam jets separated by small 
transverse coordinates too small to be resolved by the transverse momenta involved.
While the total average number of Glauber participant 
nucleons that interact with a projectile at impact parameter ${\bf b}$
is determined by the area of the inelastic cross section $\sigma_{in}(s)\sim $ few fm$^2$ as
  $N_T({\bf b})=\sigma_{in}(s) \int dz \rho_T(z,{\bf b})$, 
for moderate momentum transfers with $k_T\sim Q_s\sim$ 1-2 GeV  \br 
the target participant antennas naturally group event by event 
into $M\le N_T$ resolved clusters separated in the transverse plane  by sub-nucleon distances $1/k_T \sim 0.2$ fm,  
similar to the CGC model~\cite{Lappi:2006fp} and in AdS/CFT
 shock modeling~\cite{Noronha:2014vva} of $p+A$, but here simply to transverse resolution scale of 
 multiple scattering recoil kinematics in the target frame versus the infinite momentum frame.

This  partial decoherence of the  $N_T(b)$ participating target dipoles creates non-isotropic spatial
distributions of color antennas that radiate according to the fluctuating
spatial asymmetries from even to event. Each cluster is characterized
by the number $m_a$ of target participant dipole antennas that exchange coherently $Q_a^2=m_a \mu^2$ with the 
projectile at a specific azimuthal angle $\psi_a$ controlled by the transverse geometrical distribution of the clusters.

Each recoil cluster $a=1,\cdots,M$ 
radiates coherently into a broad range of rapidities
that appears in two particle correlations as $M+1$ ``ridge'' components
with ${\bf k}$ near the cluster accumulated recoil transverse momenta 
$-{\bf Q}_a=-\sum_{i\in I_a} {\bf q}_i$ and
with ${\bf k}$ near  the projectile dipole (cluster) radiates  near the
total momentum transfer received
${\bf Q}_0=(Q_0,\psi_0)=\sum_{a=1}^M
{\bf Q}_a$. 
On an event-by-event basis $M$ and the color antenna geometry fluctuate
producing naturally $n=M+1$ and  other azimuthal harmonics
 in  two gluon  $v_n=\langle \cos(n(\phi_1-\phi_2))\rangle$.

Our goal here is to estimate  analytically  
the magnitude of the color \br  source of pQCD dynamical
azimuthal two particle correlations and its dependence of $n,k,M,N_T$.
We illustrate the results with specific analytic cluster geometric limits, including $Z_n$ symmetric and Gaussian 
random CSA. We propose a future generalization of HIJING  
that could enable more realistic testing the 
influence of anisotropic VGA \br  on the final  hadron flavor dependent azimuthal moments 
and competing  minijet and hard jet sources of anisotropies.

\section{ First order in opacity (GB) \br   and  azimuthal asymmetries $v_n$ }

The above puzzles with BES~\cite{Adamczyk:2013gw}, $D+Au$ at RHIC,
 and with $p+Pb$ at LHC motivate us to consider an alternative,
more basic, perturbative QCD source of azimuthal asymmetries.
The well known non-abelian bremsstrahlung Gunion-Bertsch (GB) 
formula~\cite{Gunion:1981qs} for the soft gluon radiation single inclusive distribution is  
\begin{eqnarray}
 \frac{dN_g^1}{d\eta d^2 {\bf k}d^2{\bf q}}  &= &  \frac{C_R \alpha_s}{\pi^2} 
\frac{\mu^2}{\pi(q^2+\mu^2)^2}
\frac{{\bf q}^2}{{\bf k}^2 ({\bf k}-{\bf q})^2}   
\;\;,  \label{GB1} \end{eqnarray}
where we characterize the 
parton scattering elastically with the cross section
$d\sigma_{0}/d^2{\bf q}=\sigma_{0} \mu^2/\pi(q^2+\mu^2)^2$ off color neutral 
target participants with a momentum transfer ${\bf q}$ 
in terms of a characteristic cold nuclear matter scale
$\mu^2\approx 0.12$ GeV$^2$ taken from fits to forward dihadron correlations
in \cite{Qiu:2004da,Neufeld:2010dz,Kang:2012kc}. Here 
$q=|{\bf q}|$  and  the produced gluon has rapidity $\eta$ and 
transverse momentum ${\bf k}$ ($k=|{\bf k}|$) in the final state.  
It is obvious from Eq.~\ref{GB1} that non-abelian gluon \br 
is preferentially emitted along two directions specified by the 
beam ``$\hat{{\bf z}}$'' axis and the transverse momentum transfer vector ${\bf q}$. 
The uniform rapidity-even, $\eta \approx \log (x E/k)$, distribution
associated with moderate ${\bf q}$ scattering is a unique feature 
of non-abelian \br in the kinematic  $k\ll x E \ll E$ range of interest 
associated with beam jets and is due to the triple gluon vertex.  
  The uniform  rapidity-even distribution is an especially
important characteristic
of non-abelian radiation. The combination of the two 
leads to a uniform rapidity ``ridge'' in the direction of the momentum
transfer ${\bf q}$ that fluctuates in both magnitude and direction
from event-to-event but measurable in two or higher gluon correlation
measurements. The rapidity-even \br ridge is of course 
kinematically limited to $\eta\in [Y_T,Y_P]$ interval between the target 
and projectile rapidities. Independent but kinematically correlated
multiple target and projectile beam jet
\br sources can also naturally 
account for the triangle boost non-invariant rapidity density
observed in $p+A$ as emphasized in Ref.\cite{Adil:2005qn}.

For scattering of color neutral dipoles considered in~\cite{Gunion:1981qs}
the Rutherford perturbative $\alpha^2/q^4$ distribution of momentum transfers
were modeled by color neutral form factors of the form $q^2(q^2+\mu^2)^{-1}$.  
For  GB radiation  the ${\bf k}={\bf q}$ singularity is also regulated by such a form factors. Therefore the color neutralization scale 
$\mu^2$ also regulates the $({\bf k}-{\bf q})^2$ singularity in Eq. 1 as well.
That $x$ and $A$ dependence of that scale arises naturally in small $x$  models based the  gluon 
saturation scale $Q_s(x,A)$~\cite{Kharzeev:2004bw,Lappi:2006fp,Dusling:2012cg}.
Our emphasis here however is to explore the general characteristics
of $g+g\rightarrow g $ from the perturbative QCD perspectives
that allows us to derive analytically many of the observed remarkably 
simple scaling relations between $2\ell$ azimuthal harmonic cumulants,
$v_{n}(2\ell)$, as a basic coherent state semi-classical wave interference effect without invoking hydrodynamic local equilibrium assumptions.   

The screened single inclusive GB perturbative gluon distribution is
\begin{eqnarray}
 \frac{dN_g^{(1)}}{d\eta d^2 {\bf k} d^2{\bf q} } & \equiv & f(\eta,{\bf k},{\bf q} ) \nonumber\\
&=& 
\frac{C_R \alpha_s}{\pi^2k^2} 
 \frac{\mu^2q^2}{\pi({q}^2+\mu^2)^2}\frac{P_\eta}{({\bf k}-{\bf q})^2+\mu^2}  \\
 \; &   \equiv &      
 \frac{F\;P}{ A - \cos(\phi-\psi)} 
\end{eqnarray}
where $\phi$ is the azimuthal angle of ${\bf k}$ and $\psi$ is the azimuthal angle of ${\bf q}$ and abbreviations
\begin{eqnarray}
A\equiv A_{kq}&\equiv & (k^2+q^2+\mu^2)/(2 k \;q) \ge 1 \\
F\equiv F_{kq} &\equiv & \frac{C_R \alpha_s}{\pi^2k^2}
\frac{\mu^2q^2}{\pi({q}^2+\mu^2)^2}\frac{1}{2 k q}  \label{Fkq}\\
P\equiv P_{\eta}&\equiv& (1-e^{Y_T -\eta})^{n_f} (1-e^{\eta-Y_P})^{n_f}
\;\; ,\label{defs1} \end{eqnarray}
were introduced  a kinematic rapidity envelope factor $P_\eta$ 
corresponding to approximately uniform rapidity dependence 
of the non-abelian \br\cite{Gunion:1981qs} regulated
with  $(1-|x_F|)^{n_f}$ kinematic spectator
power counting ~\cite{Brodsky:1977bu, Kharzeev:2004bw}.
Note $n_f=2 n_{spec}-1 \sim 4$ for gluon production
 from the scattering of two color neutral dipoles
 in the large $|x_F|\rightarrow 1$ limit.
The $P_\eta$ rapidity envelopes 
can be used to build up 
multi beam jet boost non-invariant triangular $dN_{pA}/d\eta$ 
as in the  BGK~\cite{Brodsky:1977de} model and also to model the intrinsic
 boost non-invariance of $dN_{AA}/d\eta d{\bf x}_\perp$  even in symmetric A+A collisions as with 
HIJING~\cite{Adil:2005qn}.

\begin{figure}[!tb]
\centerline{\includegraphics[width=3.35in]{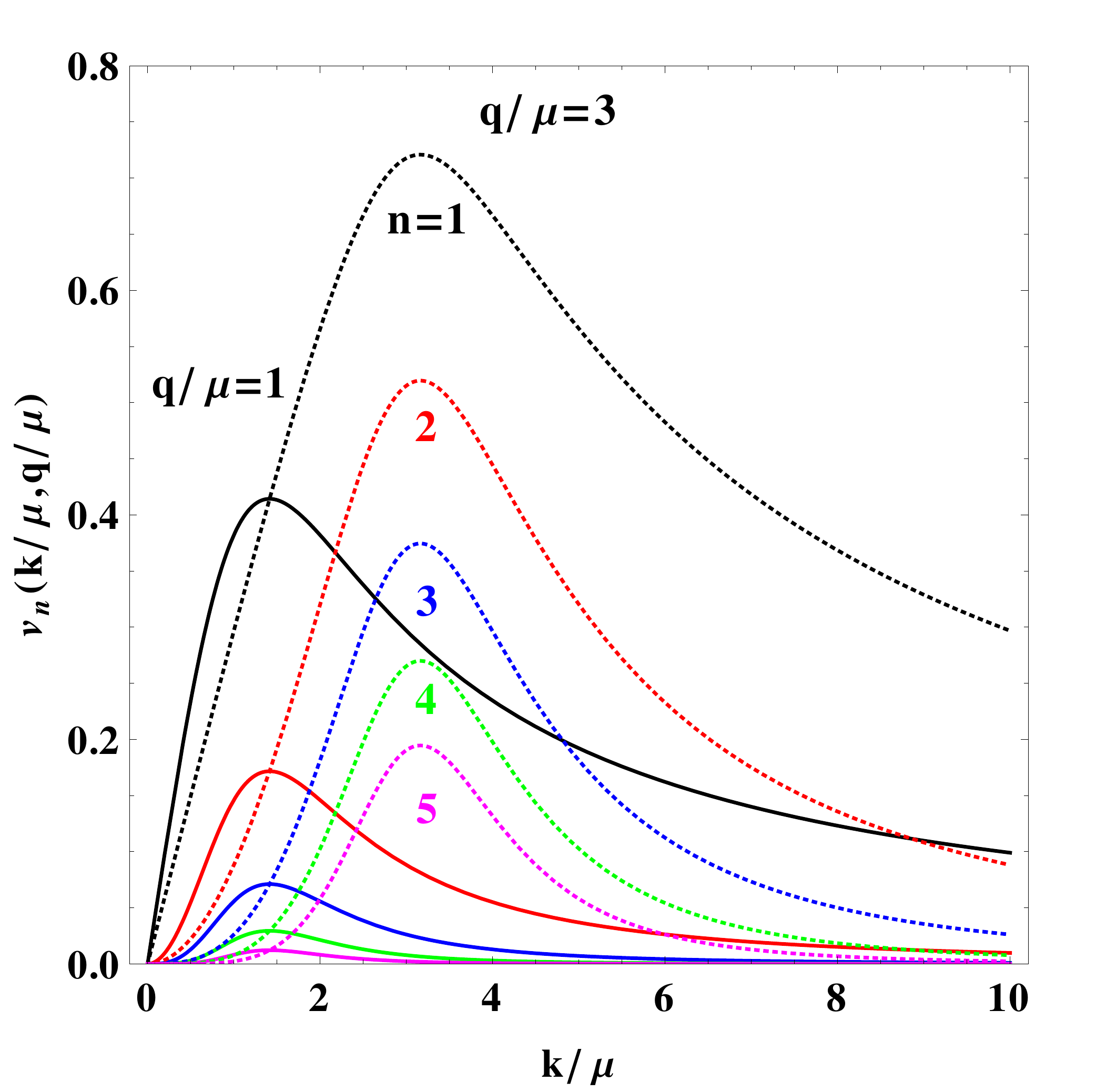}}
\caption{(Color online) Single GB beam jet \br azimuthal
Fourier moments, $ v_n^{GB}(k/\mu,q/\mu)$ 
from Eq.~(\protect\ref{vnGBscaleq}) are shown versus $k/\mu$ for $n=1-5$
for $q/\mu=1 (3)$ solid(dashed).}
\label{fig-2new}
\end{figure}
\begin{figure}[!tb]
\centerline{\includegraphics[width=3.35in]{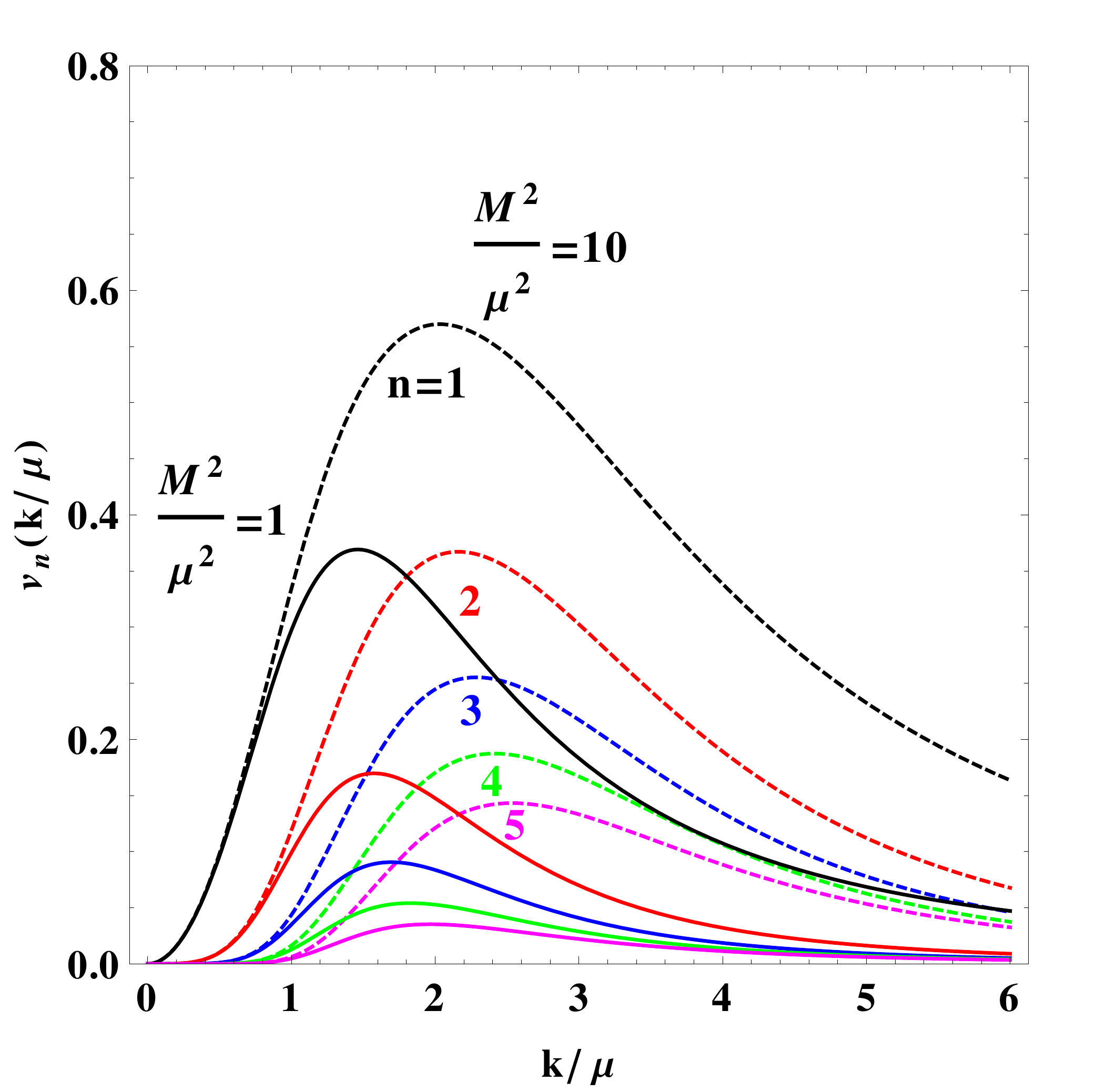}}
\caption{(Color online) Single GB beam jet \br azimuthal
Fourier moments, $ \langle v_n^{GB}(k/\mu)\rangle$, averaged over $q$ with
$M^2/(q^2+M^2)^2$  are shown versus $k/\mu$ for $n=1-5$
for $(M/\mu)^2=1\;(10)$  solid(dashed).}
\label{fig-3new}
\end{figure}
\begin{figure}[!tbh]
\centerline{\includegraphics[width=3.35in]{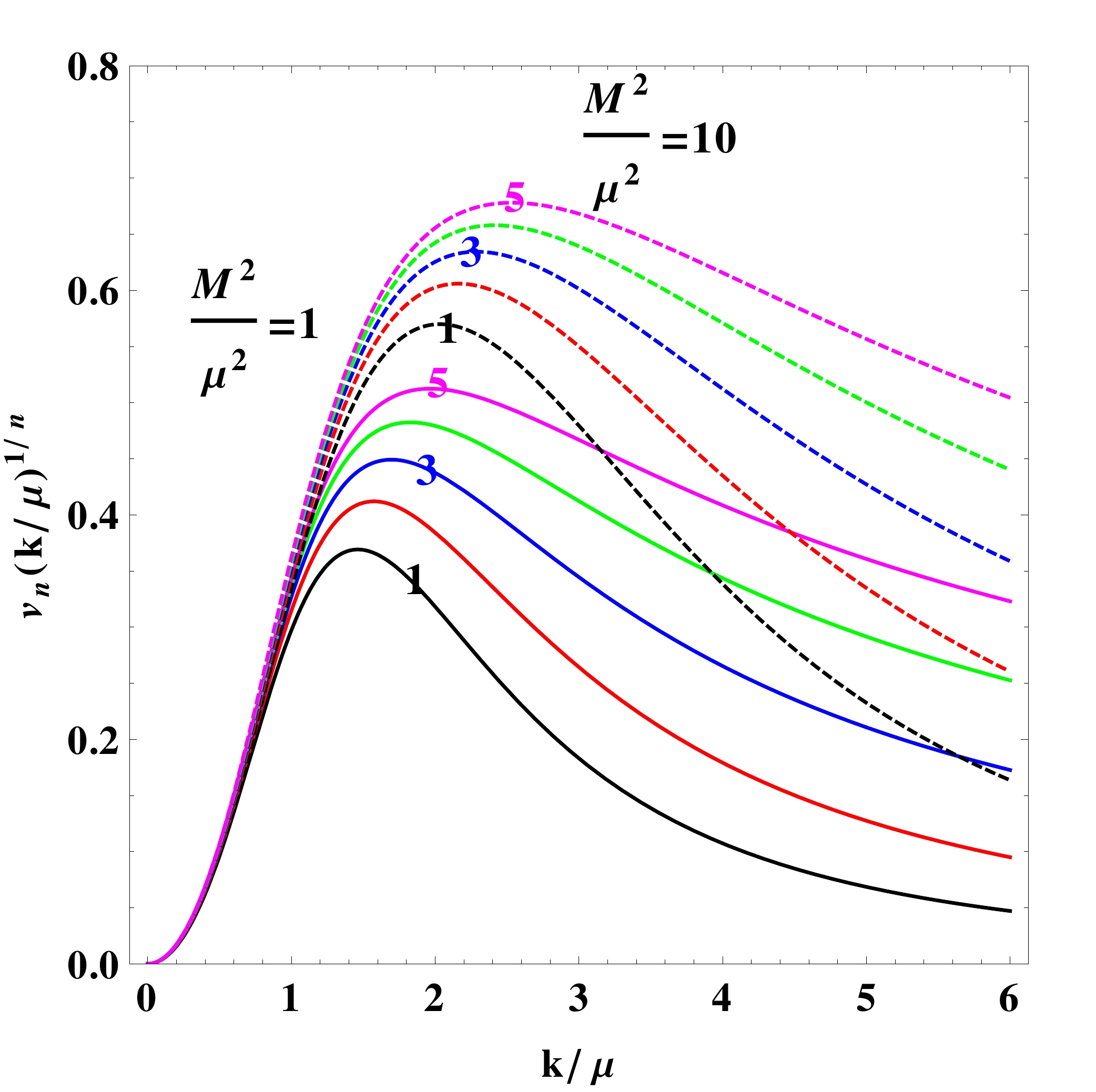}}
\caption{(Color online) Ideal $1/n$ power scaling
of $q$ averaged $ \langle v_n^{GB}(k/\mu)\rangle^{1/n}$  
with $k\stackrel{<}{\sim} M$  (see Eq.\ref{gbsc}) breaks down at higher $k$ because in the $M=0$ limit of non-abelian {\protect \br}
 limits $k\le q$ (see Eq.~(\ref{fig-4})).
}
\label{fig-4new}
\end{figure}

The single gluon azimuthal moments, $v_n=v_n\{1\}$ in cumulant notation, 
from a single GB color antenna defined by
 the  momentum transfer ${\bf q}=(q,\psi)$ with 
azimuthal angle $\psi$ are defined by
\begin{eqnarray}
\hspace{-0.2in} v_n^{GB}(k,q,\psi) f_0(k,q) 
&=& F\;P\, \int\frac{ d\phi}{2\pi} \frac{\cos(n \phi)}{ A - \cos(\phi-\psi)}
\nonumber  \\
\; &=& F \;P\,  {\rm Re}\; \oint_{|z|=1}
 \frac{dz}{2\pi i} \frac{ (-2 e^{i n \psi}) z^n}{(z^2 - 2 A z +1)}
\nonumber  \\
\; &=& F\;P\; {\rm Re}\;  \frac{ 2(e^{i\psi}\; z_-)^n}{z_+-z_-}
\; \; , \label{vn1}
\end{eqnarray}
where we defined $z\equiv \exp(i(\phi-\psi))$, so that $d\phi=-i dz/z$ and
$\cos(\phi-\psi)=(z+1/z)/2$. Note that there are two simple real 
poles $z_\pm=A\pm\sqrt{A^2-1}$. Since $A\ge 1$, only $z_-$ contributes 
to the unit contour integral, resulting in the final analytic expression above.
Note that the azimuthal averaged 
single gluon inclusive ($n=0$) \br distribution with $v_0=1$ is then 
\beqar
f_0&=& 2F\; P/(z_+-z_-)
= F_{kq}P_\eta/(A_{kq}^2-1)^{1/2}\nonumber\\ 
&\propto& {dN}/{d\eta dk^2dq^2}
\; \; . \eeqar{f0}
This has a collinear divergence at $k=q$ in the  $\mu=0$ limit in addition to 
the usual abelian beam axis $1/k^2$ divergence. The first is regulated by the
color neutral dipole form factor in the GB model.

The azimuthal Fourier moments are however finite in Eq.~(\ref{vn1}) even in the
case of vanishing $\mu$  and 
depend analytically on $n$ and $A$  via 
\begin{eqnarray}
v_1^{GB}(k,q,\psi) &=& \cos[\psi] (A_{kq}-\sqrt{A^2_{kq}-1}) \label{v1}\\
   {\displaystyle \lim_{\mu \rightarrow 0}}v_1^{GB}(k,q,0) &=& (k/q)\; \theta(q-k)\\
v_n^{GB}(k,q,\psi) &=& \cos[n \psi] \;(v_1^{GB}(k,q,0))^n \label{gbsc}\\
 {\displaystyle \lim_{\mu \rightarrow 0}} v_n^{GB}(k,q,0) &=& (k/q)^n \; \theta(q-k) 
\; \; .\label{vnGBscaleq}\end{eqnarray}
Note that  in the $\mu=0$ limit,  all $v_n\rightarrow 1$ 
reach unity at $k=q$ but vanish for $k>q$. For finite $\mu >0$, all 
moments maximizing at $k^2=k_*^2=q^2+\mu^2$
 with  $v_n(k_*) = (\sqrt(1+\mu^2/q^2)-\mu/q)^n$. Figure~\ref{fig-2new} illustrates the magnitude of 
GB  $v_n(k/\mu,q/\mu)$ moments as a function of $k/\mu$ for $n=1,\cdots,5$ and
 two different $q/\mu=1, 3$. 
 
Note the remarkable  power law scaling with $n$ (for fixed $k,q.\psi$)
 of the azimuthal moments
of gluon \br from  a single GB color antenna:
\begin{eqnarray}
[v_n^{GB}(k,q,0)]^{1/n} &=& \;[v_m^{GB}(k,q,0)]^{1/m}
\; \; ,\label{vnscaling}\end{eqnarray}
that is similar to the scaling observed by ALICE, CMS and 
ATLAS~\cite{Aamodt:2011by, Chatrchyan:2012wg,Aad:2012gla} at LHC
at least for the higher $n\ge 3$ moments 
dominated by purely geometric fluctuations. 
This scaling is of course  not expected to hold  
perfectly  for ensemble $q$ averaged
ratios of $\cos(n\Delta \phi)$ of di-hadron inclusive rates.
One of our aims below is to test the survival
 of the above ideal scaling in Eq.~(\ref{vnscaling}) to ensembles averages in
two gluon inclusive processes.

However, note that by rotation invariance 
all harmonics $n>0$ vanish for {\em single} inclusive
GB antennas when averaged over the momentum transfer azimuthal angle $\psi$.
We show below in section V that the finite {\em rms} fluctuating harmonics 
of {\em two particle}  inclusive
$(\langle \cos(n \Delta\phi) \rangle)^{1/2}$ survive with 
similar magnitude and $k$ dependence as in Figs.~1,2.

In Fig. \ref{fig-4new} we see that the simple fixed $q$ power law scaling of
Eqs.~(\ref{gbsc},\ref{vnscaling}) holds for $k/M <1$ but gradually
breaks down at higher $k>M$ when  ensemble averaged over $q^2$ in
$\langle f_n(k)\rangle$.

\section{All Orders in Opacity VGB Generalization of Gunion-Bertsch radiation}

A recursive reaction operator method was originally developed in GLV~\cite{Gyulassy:2000er,Gyulassy:2003mc} to compute final-state 
multiple collision-induced gluon \br and elastic collisional energy loss~\cite{Gyulassy:2002yv} to all orders in opacity
for applications to jet quenching. Extensions of the method to final state  heavy quarks jet energy loss was given 
in~\cite{Djordjevic:2003zk,Wicks:2005gt}.

Vitev further extended the reaction operator method to compute non-abelian energy loss in cold nuclear matter in Ref.~\cite{Vitev:2007ve}. 
In addition to  Final-State (FS) \br, Vitev solved the cold matter Initial-State (IS) \br problem to all orders in opacity and also 
the generalization of the first order in 
opacity Gunion-Bertsch~\cite{Gunion:1981qs}  non-abelian bremsstrahlung problem to all orders in opacity for asymptotic 
($t_0\rightarrow -\infty, t_f \rightarrow +\infty$)  boundary condition. We refer here to the Vitev all-order in opacity generalized 
GB radiation solution as VGB.

In~\cite{Vitev:2007ve} the VGB solution was regarded to be of mainly 
academic interest, since the focus there was on induced initial state IS and final state FS 
gluon bremsstrahlung associated with hard processes in $p+A$~\cite{Qiu:2004da,Neufeld:2010dz,Kang:2012kc}. 
In this paper,  we focus entirely on the application of the VGB solution to low to moderate transverse momentum 
$k<$~few GeV  gluon radiation from  multiple beam jets in the same spirit
as in GB~\cite{Gunion:1981qs}, where the aim was to understand  the general qualitative characteristics 
of inelastic high energy single inclusive processes from low order perturbative QCD perspective.

Our aim here is to calculate azimuthal asymmetry moments, $v_n(\eta,{\bf k})$, arising from basic 
  perturbative QCD bremsstrahlung effects  in high energy p+A interactions. The physical picture approximates  
p+A scattering as  the scattering of an incoming  color
dipole at an impact parameter, ${\bf b} $,  of high (positive) rapidity $Y_P\gg 1 $ with 
 $N_T^{part}~\sim A^{1/3}$ nuclear target participant nucleons with high (negative) $Y_T \ll - 1$ in the CM.
The target participant dipoles at a fixed transverse coordinate ${\bf R}$ 
are separated  longitudinal separations $\Delta z_i = z_i -z_{i-1}\sim $ fm in the cold nucleus target rest frame. 
However they act   coherently when emitting gluons near mid rapidity due to Lorentz contraction 
in the CM and long formation time of gluons $ \sim [2 \cosh (\eta)]/k$ in the lab frame.

However, the target participants are distributed in the transverse direction by
 transverse separations $R_{ij}=|{\bf R}_i -{\bf R}_j  |\stackrel{<}{\sim} \sqrt{\sigma_{in}/\pi}\sim $~fm  
which can be resolved for $k>1$ GeV.  This leads incoherent groups  
of target nucleons that radiate mid-rapidity gluons with $|\eta|<1$, $k>1$~GeV gluons coherently. 
We propose in section IV below  a simple percolation model to estimate  
the partially coherent target recoil \br.  However, we concentrate in this section  on the coherent projectile \br contribution.

The complete all orders in opacity, 
$\chi\equiv\chi({\bf b})=\int dz \sigma_g(z)\rho(z,{\bf b})$,
 VGB solution derived by Vitev in \cite{Vitev:2007ve} is 
\begin{widetext} 
\beqar
 \frac{dN^{VBG}}{d\eta d^2 {\bf k} } &=& \sum_{n=1}^{\infty} 
\frac{dN_n^{VBG}}{d\eta d^2 {\bf k} }= 
\frac{C_R \alpha_s}{\pi^2}  \sum_{n=1}^{\infty} \left[ \prod_{i = 1}^n \int
\frac{d \Delta z_i}{\lambda_g(z_i)}  \right] 
\left[ \prod_{j=1}^n \int d^2 {\bf q}_j \left(v_j^2({\bf q}_j)
-  \delta^2 ({\bf q}_j) \right)    \right] \nonumber \\ 
&& \times  \;  {\bf B}^b_{21} \cdot 
\left[ {\bf B}^n_{21}
 + 2 \sum_{i=2}^n {\bf B}^n_{(i+1)i} 
\cos \left( \sum_{j=2}^i \omega_{jn} \Delta z_j  
\right) \right] \;,
\eeqar{BGfull}
\end{widetext}
where the transverse vector ``antenna'' amplitudes 
${\bf B}^n_{jk}$   are defined in terms of differences 
between``cascade'' vector amplitudes ${\bf C}_{jn}$ as
\begin{eqnarray}
{\bf B}^n_{jk} &=&  {\bf C}_{jn} -{\bf C}_{kn} \\
{\bf C}_{jn}&=&  \frac{{\bf k} - {\bf q}_j - \cdots - {\bf q}_n}
{({\bf k} - {\bf q}_j - \cdots - {\bf q}_n)^2}  =   \frac{{\bf k} - {\bf Q}_{jn} }
{({\bf k} -{\bf Q}_{jn} )^2}\;\; .
\label{defsC}
\end{eqnarray}
Indices $j,k,n$ here  keep track of combinations of 
non-vanishing momentum transfers
${\bf q}_i$  from direct versus virtual diagrams contributing at a 
given opacity order $n$ of the opacity expansion. The partial summed
 momentum transfers are ${\bf Q}_{jn}=\sum_{i=j}^n{\bf q}_{i}$ being the singular directions of non-abelian bremsstrahlung that 
also control the inverse formation times 
\beq
\omega_{jn}= \frac{({\bf k}-{\bf Q}_{jn})^2}{2E_g}\, .
\eeq{omeg}
Here $E_g= x E_P \ll E_P$ is the energy of the gluon in a frame where the 
energy of the proton projectile is assumed to be large $E_P \gg m_n$.

There are two simple limits depending on the kinematic range of interest.
In the fully coherent limit, where $n \omega_{jn}\lambda_g \ll 1$,
we can approximate all the cosines by unity. This is the limit we are interested
in for our present applications to mid-rapidity multi-particle production
not too close to projectile and target fragmentation regions, i.e  
$Y_T+1  < \eta< Y_P-1$. 

The target scattering centers are ordered in this VGB problem as
$z_0=-\infty< z_1< \cdots<z_n< z_f=+\infty$ with $\Delta z_i=(z_i-z_{i-1})$ for $i\ge 2$.    
$\sigma_g(z)\rho(z,{\bf b})$ is the local inverse mean free path of a gluon
the nuclear target at position z 
impact parameter ${\bf b}$ in the target rest frame. 
The ${v}^2({\bf q}_j)=\frac{d \sigma_{el}(z_j) }{d^2 {\bf q}_j} $ denote
normalized distributions of transverse momentum transfers at scattering center $z_j$.

In the coherent scattering limit of relevance to  near mid-rapidity
radiation and neglecting possible $z$ dependence of the screening scale $\mu$
of the normalized distribution $v^2({\bf q})$, we can write more explicitly
at impact parameter ${\bf b}$

\begin{widetext}
\beqar
\frac{dN^{VGB}_{coh}}{d\eta d^2 {\bf k} } &=& 
\frac{C_R \alpha_s}{\pi^2}\sum_{n=1}^\infty
\left[ \prod_{i = 1}^n \int
{d \Delta z_i}\; { \sigma_{el}(z_i) \rho(z_i,{\bf b})}  \right] 
\left[ \prod_{j=1}^n \int d^2 {\bf q}_j    
 \left( \frac{1}{\sigma_{el}} 
\frac{d \sigma_{el}}{d^2 {\bf q}_j}   
-  \delta^2 ({\bf q}_j) \right)    \right]  \nonumber \\ 
&\;& \hspace*{-0.5cm} \times  \; 
\left( \frac{{\bf k} - {\bf q}_2 - \cdots - {\bf q}_n}
{({\bf k} - {\bf q}_2 - \cdots - {\bf q}_n)^2}  - 
 \frac{{\bf k} - {\bf q}_1 - \cdots - {\bf q}_n}
{({\bf k} - {\bf q}_1 - \cdots - {\bf q}_n)^2}     \right)
\cdot 
\left[ \left( \frac{{\bf k} - {\bf q}_2 - \cdots - {\bf q}_n}
{({\bf k} - {\bf q}_2 - \cdots - {\bf q}_n)^2}  - 
 \frac{{\bf k} - {\bf q}_1 - \cdots - {\bf q}_n}
{({\bf k} - {\bf q}_1 - \cdots - {\bf q}_n)^2}     \right) 
 \right. \nonumber \\
&\;& \hspace*{-0.5cm} \left.  + 2 \sum_{i=2}^n 
\left( \frac{ {\bf k} - {\bf q}_{i+1} - \cdots - {\bf q}_n }
{ ({\bf k} - {\bf q}_{i+1} - \cdots - {\bf q}_n )^2 }  - 
 \frac{ {\bf k} - {\bf q}_i - \cdots - {\bf q}_n }
{ ( {\bf k} - {\bf q}_i - \cdots - {\bf q}_n)^2 }     \right) 
\right] \, .
\eeqar{VGBf}
\end{widetext}

In order extract the the physical interpretation of
 the above complete but unwieldy expression, 
we derive in the Appendix A the 
linked cluster theorem version of Eq.~\ref{VGBf} to be 
\beq 
 dN^{VGB}_{coh}({\bf k})= \sum_{n=1}^\infty 
 \int d^2{\bf Q} \; P^{el}_n({\bf Q})\; dN^{GB}({\bf k},{\bf Q}) 
\;\; ,\eeq{linkclusterth}
where $P^{el}_n({\bf Q})$ is the probability 
density 
that after $n$ elastic scatterings the cumulative total momentum transfer is ${\bf Q}$, 
 \beqar 
  P^{el}_n({\bf Q})&=& \exp[{-\chi}]\frac{\chi^n}{n!} \int  \left\{ \prod_{j=1}^n \frac{d^2{\bf q}_j}{\sigma_{el}}
\frac{d \sigma_{el}}{d^2 {\bf q}_j}   
 \right\} \nonumber \\ 
&\;&\hspace{0.5in} \times \delta^2({\bf Q} - ({\bf q}_1+\cdots +{\bf q}_n))
\;\; ,\eeqar{elastQ}
that is independent of the azimuthal direction $\psi$ of ${\bf Q}$
by rotation invariance. This distribution also arose naturally in 
 the reaction operator derivation of the link cluster theorem 
for multiple elastic scattering in Ref.~\cite{Gyulassy:2002yv}.

Eq.~(\ref{linkclusterth}) is clearly the intuitive factorization limit
where at each order only the total accumulated momentum transfer, ${\bf Q}$,
controls the azimuthal and momentum transfer dependence of 
the \br distribution. 

\begin{figure}[!tb]
\centerline{\includegraphics[width=3.35in]{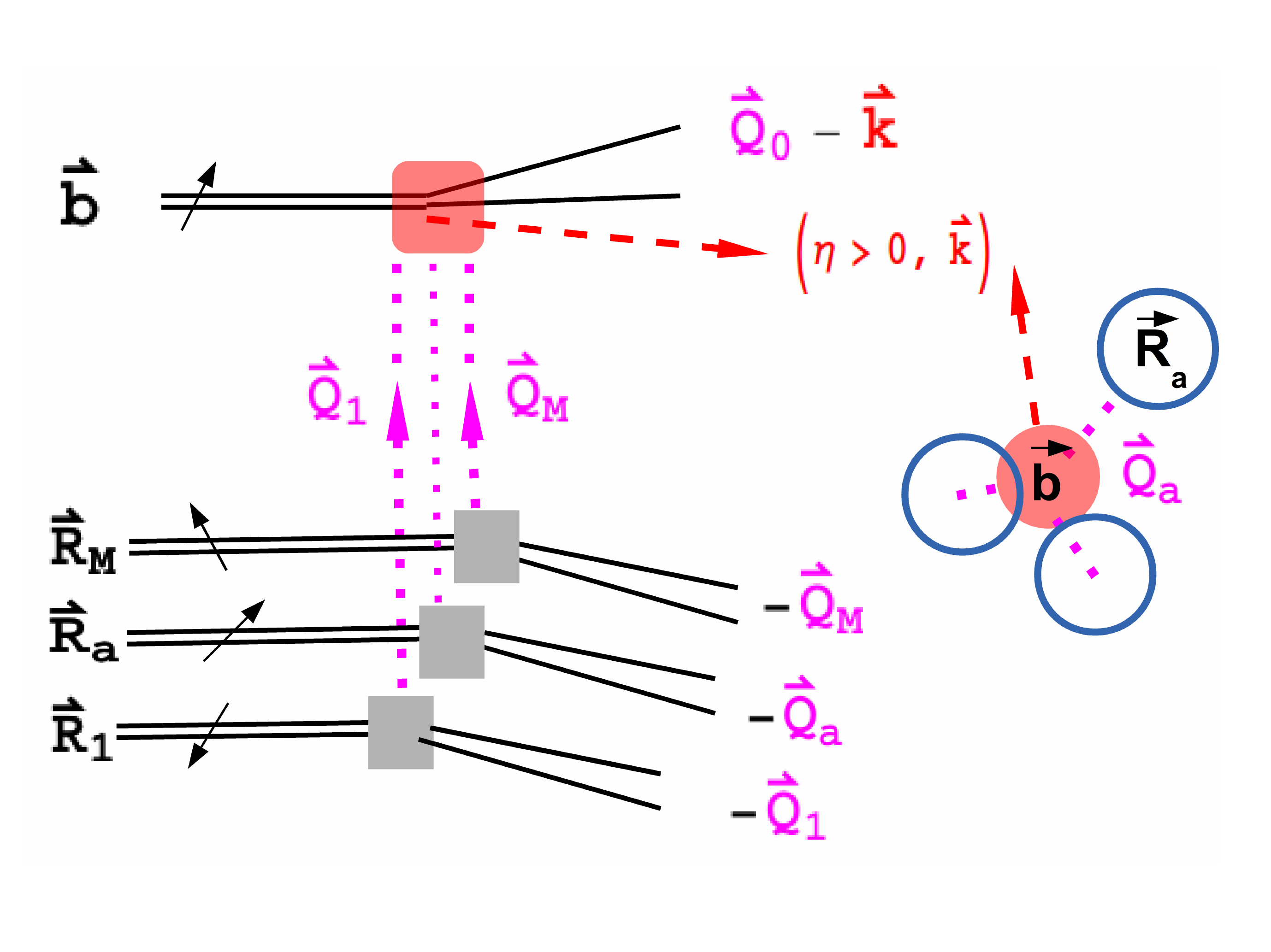}}
\caption{(Color online) Schematic diagram corresponding 
to coherent \br from the projectile dipole from
Eqs.~({\protect \ref{linkclusterth},\ref{elastQ}}).  At opacity order $n$ 
the azimuthal distribution
is enhanced for transverse momenta ${\bf k}$ near the total accumulated
momentum transfer
${\bf Q}_0\equiv {\bf Q}_{1n}=\sum_{a} {\bf Q}_a$ where $a=1,\cdots, M$
groups of recoiling target dipoles.}
\label{fig-1}
\end{figure}
\begin{figure}[!tb]
\centerline{\includegraphics[width=3.35in]{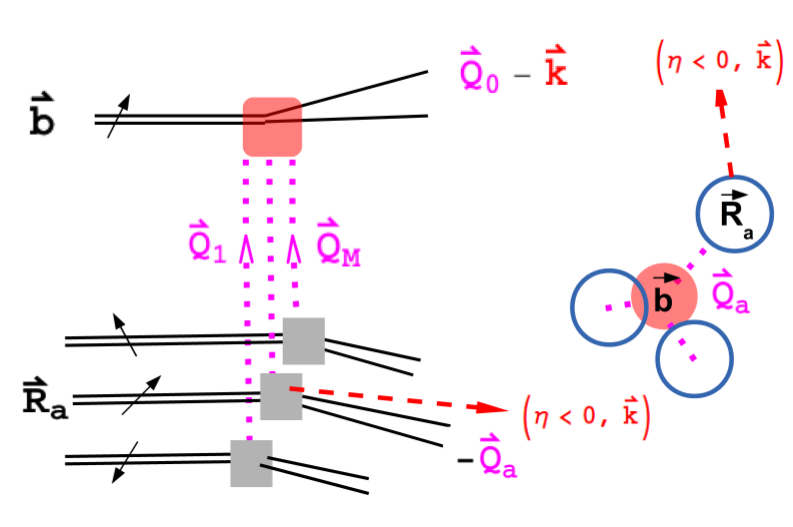}}
\caption{(Color online) Schematic diagram corresponding 
to partial coherent  gluon bremsstrahlung from 
Eqs.~({\protect \ref{GBT}}).  At opacity order $n$ 
the azimuthal distribution
is enhanced in transverse momenta ${ \bf k}$ near the recoil momentum transfers
$-{\bf Q}_a$ where $a=1,\cdots, M$
labels incoherent target groups of color dipoles fragmenting toward the negative
rapidity region.}
\label{fig-2}
\end{figure}

By rotation invariance $dN^{GB}({\bf k},{\bf Q}) = dN^{GB}(k,Q,\phi-\psi)$
can only depend on the k and Q azimuthal angles through their difference.
After integrating over $\psi$, the azimuthal angle of ${\bf Q}$, then of course $dN^{VGA}$ 
cannot depend on the azimuthal angle $\phi$ of ${\bf k}$.
Therefore,  it is obvious that at the single inclusive level
all $v_n=0$ vanish for $n>0$. To observe the intrinsic fluctuating azimuthal asymmetries event-by-event
we turn to two particle correlations to extract non-vanishing
second moments like $\langle\cos(n(\phi_1-\phi_2))\rangle$. 
First we discuss the \br contribution from recoil target participants. 

\section{\br \   from Recoiling Target Participants}

Incoherent groups of transversely overlapping recoiling target dipoles
radiate gluon bremsstrahlung dominantly into the negative 
rapidity $\eta<0$ hemisphere,  as illustrated in Fig.~\ref{fig-2}.  In a given event when a projectile
nucleon penetrates through a target nucleus $A$  at impact parameter
${\bf b}$, the projectile  nucleon moving with positive rapidity $Y_P>0$
is approximated as in Ref.~\cite{Gunion:1981qs}  by a color dipole with a separation  ${\bf d}_0=\hat{n}_0/\mu_0$ .
The $A$  target nucleons moving toward negative rapidities, $Y_T<0$, are however 
distributed with transverse  coordinates ${\bf R}_i$, according to a Glauber nuclear profile
distribution $T_A({\bf R}_i)=\int dz \rho_A(z,{\bf R}_i)$ over a large area $\pi A^{2/3}$ fm$^2$ scale. Each target 
nucleon dipole is assume to have a separation
${\bf d}_i=\hat{n}_i/\mu_i$. Projectile target   dipole-dipole interactions
with low transverse momentum transfer ${\bf q}_i < \mu_i$ are suppressed by
dipole form factors approximated by $q_i^2/(q_i^2+\mu_i^2)$. Therefore,
the projectile interacts dominantly with only  nearby target dipoles in the transverse plane with 
$({\bf R}_i-{\bf b})^2 \stackrel{<}{\sim}  \pi \alpha^2(d_0 +d_i)^2/4
\sim \sigma_{in}$. This leads to a fluctuating number $n$ 
of target participants with probability $P_n=e^{-\chi}\chi^n/n!$ that follows
also from the  GLV opacity 
expansion~\cite{Gyulassy:2000er, Gyulassy:2002yv,Vitev:2007ve}.   

For a given target participant number, $n$, the target dipoles naturally cluster
near the projectile  impact parameter ${\bf{b}}$ as illustrated in Figs.~(\ref{fig-1},\ref{fig-2}). 
In a specific event,  there are in general   $1\le M\le n$ 
overlapping clusters that radiate coherently toward 
the negative rapidity $\eta<0$ hemisphere
 as illustrated in  Fig.(\ref{fig-2}). The distribution of the number $M$ of
recoiling coherent groups depends on $n, {\bf k}$, and the momentum exchanges ${\bf q}_i$ 
with the projectile that build up to the total exchange to the projectile 
\beq
{\bf Q}_P=\sum_{a=1}^M{\bf Q}_a=\sum_{a=1}^M(\sum_{i\in I_a} {\bf q}_{i})\; ,\eeq{QP}
where $I_a$ is a particular subset of the $n$ indices $i\in[1,n]=\sum_a I_a$ 
that the emitted gluon with transverse wavenumber $k$ (and generally $\eta <0$) 
cannot resolve, and  ${\bf Q}_a=\sum_{i\in I_a} {\bf q}_{i}$ is the 
contribution from  group $I_a$ to the total momentum transfer 
to the projectile.

A simple percolation model for identifying clusters of coherently recoiling
target groups of dipoles  is to require that all members in a cluster
have separation ${\bf R}_{ij}={\bf R}_i-{\bf R}_j$ in the transverse plane 
in modulus less than the produced gluon transverse momentum 
resolution scale, i.e. 
\beq
R_{ij}\stackrel{<}{\sim} d(k) = \frac{c}{k}
\;\;,\eeq{percol}
where $c\sim 1$ is of order unity.
If $i\in I_a$ and $j\in I_a$ as well as $j\in I_b$, then $j$ is added to $I_a$ if its $\langle d_{ij}\rangle_{i\in I_a} < \langle d_{ij}\rangle_{i\in I_b}$.
The M clusters are percolation groups in the above sense. Of course 
many other variants of transverse clustering algorithms exist. For our purpose of illustrating analytically 
dynamical sources $v_n$ in p+A compared to peripheral $A+A$ it suffices to study the dependence of $v_n$ 
on the number of independent recoil antennas $M$ with $<n>=N$ fixed by 
Glauber participant geometry. In future applications via Monte Carlo
generators such as HIJING~\cite{Wang:1991hta} the sensitivity of results to more realistic multi beam jet geometric fluctuations 
can be studied.  Note that independent  target participant beam jet clusters are cylindrical cuts
 into the target frame near the impact parameter ${\bf b}$ 
with diameters $\sim 1/k$. We expect typically $M\sim 2-4$ independent
recoil clusters  even for the most central $p+A$ collisions,  as illustrated in  Figs.~(\ref{fig-1},\ref{fig-2}). This picture is similar to the CGC model
picture except that no classical longitudinal fields are assumed in our entirely perturbative QCD dynamical \br approach here.  

In a given event, recoil bremsstrahlung  contribution  to the single inclusive gluon distribution from  $M$
coherently acting but transversely resolvable target antenna clusters  is given by 
\beqar
dN^{M,N}_{T}(\eta,{\bf k};\{{\bf q}_{j}\})\equiv  
\sum_{a=1}^M dN^{GB}({\bf k},-{\bf Q}_{a}) P_a(\eta)\;\; , 
\eeqar{GBT}
where $P_a(\eta)$ specifies different  rapidity profile functions for each cluster required 
to produce the characteristic BGK~\cite{Brodsky:1977de}
 boost non-invariant triangular enhancement 
of the rapidity density, $(dN_{pA}/d\eta)/(dN_{pp}/d\eta)$,
growing toward the value $<n>=N$ near the target rapidity $Y_T$
and dropping  toward unity near  the projectile rapidity $Y_p$. 
 
In the special doubly coherent projectile and target limit with $M=1$,
$dN_T^{1,N}$ reduces to 
\beqar
dN^{1,N}_{T}(\eta,{\bf k};\{{\bf q}_{i}\}_n)\equiv  
dN^{GB}({\bf k},\;-{\bf Q}_P)P_T(\eta)\; ,
\eeqar{b}
with $P_T(\eta)=\sum_a P_a(\eta)$. Note that in the high energy small $ x^-\propto \exp[Y_T-\eta]$  gluon saturation dynamics 
correlates $ {\bf Q}_P$ with rapidity $\eta$ instead of the simple factorization assumed in Eq.~(\ref{b}).
In our simple perturbative dipole picture this correlation can be implemented parametrically by taking $\mu_i(\eta)\propto Q_s(\eta,A)$
\cite{Kharzeev:2004bw, Lappi:2006fp, Dusling:2013oia}.

The fully coherent  projectile bremsstrahlung contribution is
\beqar
dN^{M,N}_{P}(\eta,{\bf k};\{{\bf q}_{i}\})\equiv  
dN^{GB}({\bf k},\;+{\bf Q}_P) P_0(\eta).
\eeqar{Pnco}
For $p+p$ scattering with $M=N=1$, the sum reduces in the  CM to 
\beq
dN_{pp}=  dN^{GB}({\bf k},\;+{\bf Q}_P)P_P(\eta)+  dN^{GB}({\bf k},\;-{\bf Q}_P)P_P(-\eta) \; \; ,
\eeq{dn.}
which is symmetric with respect to changing the 
sign of the total momentum transfer, ${\bf Q}_P$, 
as well as to reflecting $\eta$.

In the more general partially coherent target case with $1< M \le N $ 
independent clusters of dipole antennas,
the total single inclusive  radiation distribution in mode $({\bf k}_1,\eta_1)$
is 
\beqar
 dN^{M,N}&=& dN^{N}_{P}(\eta,{\bf k}_1 ; {\bf Q}_P)
+dN^{M,N}_{T}(\eta,{\bf k}_1;\{{\bf Q}_{a}\})\hspace{0.5in}\nonumber \\
&=&   
\sum_{a=0}^{M}  
\frac{ B_{1a}}{ 
({\bf k}_1+{\bf Q}_a)^2+\mu_a^2 } 
\;\; , 
\eeqar{GBPTM}
where we defined 
${\bf Q}_0 \equiv -{\bf Q}_P= -\sum_a {\bf Q}_a$ to be able to 
include the projectile contribution  into the summation over target clusters. 
The numerator factor $B_{ia}$ is defined 
using Eqs.~(\ref{Fkq},\ref{defs1}) to be
\beq
B_{ia}\equiv F_{k_i,Q_a}\; P_a(\eta_i)
\;\; . \eeq{bia}

For a fixed set of ${\bf Q}_a=(Q_a,\psi_a)$ of independent recoil momenta, 
the single gluon inclusive azimuthal Fourier moments 
$\langle \cos(n\phi)\rangle$ 
are given by linear combinations of $v_n^{GB}(k_1, Q_a)\cos(n\psi_a)$
from Eqs.~ (\ref{vn1})-(\ref{vnGBscaleq}). However,  since all the terms in the sum
contribute with one of $M+1$
$\cos( n \psi_a)$ factors, averaging over rotations $\psi_a\rightarrow \psi_a+\theta$  again causes all ensemble averaged
$<v_n>=0$ to vanish for $n\ge 1$. In order to extract information
about the relative {\em fluctuating} $v_n$, we therefore turn to 
two gluon correlations in the next section.
 
\section{ Multi gluon cumulant azimuthal harmonics, $v_n\{2\ell\}$, 
from Color Scintillation Antenna (CSA) arrays}

Multiple \br gluons are radiated over long ranges (``ridges'')
in $Y_T< \eta_i<Y_P$ from    multiple kinematically and transverse space 
correlated beam jets that form ``Color Scintillation Antenna'' (CSA) arrays that fluctuate from event to event.
Depending on the transverse space geometry, ${\bf R}_a$ and 
the transverse momentum transfers , ${\bf Q}_a$, and their distributions,
the CSA \br leads to fluctuating patterns of azimuthal correlations
among the radiated gluons.  Gluon \br from a single beam jet color dipole 
antenna builds up a ``near side'' correlations.
Kinematic recoil momentum correlated $N$  participant
target and projectile antennas, however, also 
 naturally radiate with  $k^2\sim N \mu^2$ in complex fluctuating 
azimuthal harmonic \br patterns. At much higher transverse momenta $k^2\gg M\mu^2$,
collinear factorized back-to-back hard jet production dominates over multiple beam jets 
\br and leads to very strong away side $n=1$ correlations that must be subtracted
in order  to reveal the moderate $k^2\stackrel{<}{\sim} M \mu^2$ correlations that we compute
here. We also assume that we can neglect a possibly large in magnitude 
transverse isotropic non-perturbative bulk background through appropriate
experimental mixed event subtraction schemes.  

Assuming that $M$ antenna clusters out of the $N=N_T^{part}({\bf b})$ 
target participants  radiate independently - 
i.e., assuming that each cluster in the CSA array 
    produces approximately a semi-classical
  coherent state of gluon radiation with random phase with respect to other clusters 
(see analogous partially coherent pion interferomentry formalism in Ref.\cite{Gyulassy:1979yi}) - 
the even number $2\ell$ inclusive gluons distribution  factorizes as
\begin{widetext}
\beqar 
 dN^{M}_{2\ell}(\eta_1,{\bf k}_1,\cdots, \eta_{2\ell}{\bf k}_{2\ell})
&=& \prod_{i=1}^{2\ell} \left(\sum_{a_i=0}^M \frac{B_{k_i a_i}}{A_{k_i a_i}-\cos(\phi_i+\psi_{a_i})}\right)
\;\; ,\eeqar{sca2M}
\end{widetext}
where $B_{ia}$ is defined in Eq.~(\ref{bia}) and again the summation range
includes the projectile $a=0$ contribution with ${\bf Q}_0\equiv -{\bf Q}_P$.
We emphasize that the total gluon inclusive has in addition to $dN^M_{2\ell}$
an isotropic $dN^{non.pert.}_{2\ell}$ and a highly away side correlated 
$dN^{dijet}_{2\ell}$  components that we assume can be subtracted away.
Implicitly we also assume here the greatly simplified ``local parton hadron'' duality 
hadronization prescription as in CGC models. 
Of course, in CGC saturation models the details, especially the $x, A,$ and ${\bf b}$ will differ, but it is useful to explore here the basic
consequences of this simple analytic model to get a feeling of
how much of the azimuthal fluctuation phenomenology may have its roots
in low order Low-Nussinov/Gunion-Bertsch pQCD interference phenomena.  
Quenching of $v_n$ \br harmonics due especially 
to more realistic hadronization phenomenology~\cite{Wang:1991hta, Andersson:1986gw,Sjostrand:1993yb} in the few GeV minijet scale 
will also need to be investigated
in the future.

Even with uncorrelated gluon number 
coherent state product ansatz for the multi gluon
inclusive distribution above, the even number $m=2\ell$ gluons with $({\bf k}_1,\eta_1)$ to $ ({\bf k}_m,\eta_m)$ become  
correlated through the CSA geometric and kinematic recoil correlations.

Consider, for example,  the $M=2$ case (see Appendix B) of 
 two recoiling target dipole antennas that emit  ${\bf k}_1$
preferentially near $-{\bf Q}_1=(q_1,\psi_1+\pi)$ and
 near $-{\bf Q}_2=(q_2,\psi_2+\pi)$,  at 
two different recoil azimuthal angles
$\psi_1+\pi$ and $\psi_2+\pi$,  while the projectile dipole emits  ${\bf k}_2$ preferentially
near ${\bf Q}_P ={\bf Q}_1+{\bf Q}_2$ at a third $\phi_P$ azimuthal angle. 
Such a three color antenna system then naturally leads to two particle
 triangularity $v_3\{2\} \equiv \langle \cos(3(\phi_1-\phi_2))\ne 0$ due to dynamical
correlations between ${\bf k}_1$ and ${\bf k}_2$. As we also show below in section V, 
special cases of $Z_n$ symmetric antenna arrays illustrate
``perfect'' \br leading to a
 pure $v_{n'}\{2\}=\delta_{nn'}v_{n}^{Z_n}\{2\} $ two particle harmonic.  

Consider in detail the prototype $M=1$ VGB antenna case again but 
for $2\ell$ gluon 
cumulant $n^{th}$ relative 
harmonic moments.  For a fixed ${\bf Q}$ impulse,
\begin{widetext}
\beqar
f^{M=1}_n\{2\ell\}&\equiv& \left\langle 
e^{ +i\; n \left\{\sum_{i=1}^\ell \phi_i\right\} } 
e^{-i\;n \left\{\sum_{j=\ell+1}^{2\ell} \phi_j\right\} } \right\rangle \; f^{M=1}_0\{2\ell\}
\nonumber\\
&=& 
\prod_{i=1}^\ell \left(\int\frac{d\phi_i}{2\pi}
\frac{B_{k_i Q}\;e^{+in\phi_i}}{A_{k_i Q}-\cos(\phi_i+\psi_Q)}\right) 
\prod_{j=\ell+1}^{2 \ell} 
\left(\int\frac{d\phi_j}{2\pi}
\frac{B_{k_j Q}\;e^{-i n\phi_j}}{A_{k_j Q}-\cos(\phi_j+\psi_Q)}\right) \nonumber \\
&=& 
\prod_{i=1}^\ell  \left(e^{in\psi_Q} \left(z_{k_iQ}\right)^n\;f_{0,k_i,Q}\right)
\prod_{j=\ell+1}^{2 \ell} \left( e^{- i n\psi_Q} \left(z_{k_jQ}\right)^n\;f_{0,k_j,Q}\right)
\nonumber\\
&=& f^{M=1}_0\{2\ell\}
\prod_{i=1}^{2\ell}  \left(v_1^{GB}({k_i,Q}\right)^n 
\;\; . \end{eqnarray}
\end{widetext}
Note that by construction even gluon number $f_n^M\{2\ell\}$ are $SO(2)$ 
rotation invariant about the beam axis and thus
independent 
of the random orientation, $\psi_Q$, of the reaction plane
defined by the transverse momentum transfer ${\bf Q}$.
Of course odd gluon number cumulants vanish after averaging over the reaction plane.

Here $z_{k_iQ}= A_{k_iQ}-\sqrt{A_{k_iQ}^2-1}$ 
are the poles inside the unit circle that contribute to the nth harmonics.
For odd number of gluons all harmonics vanish but for even
numbers all harmonics both even and odd are generated already
by one $M=1$ color GB \br antenna. For $M=2$, two recoiling GB antennas, 
${\bf Q}$ and $-{\bf Q}$ all odd $n=1,3,\cdots$ moments vanish by symmetry.
An $M$ odd number of antennas are needed to generate odd $n$ harmonics through  even number of gluon correlators.

In the ``mean recoil'' approximation $Q\approx \bar{Q}$,
we see that a single GB antenna satisfies the generalized power scaling
law in case that subsets of the $2\ell$ gluons
have identical momenta. Suppose  there are $1\le L\le 2\ell$
distinct momenta $K_r$  with $r=1,\cdots,L$ such 
$m_r$ of the $2\ell$ gluons have  momenta equal 
to a particular value $K_r$ such that $\sum_{r=1}^{L} m_r= 2\ell$.
In this case 
\beqar
v_n^{M=1}\{2\ell\}(k_1,\cdots,k_{2\ell};\bar{Q} )
&\approx &  \prod_{r=1}^L (v_n^{GB}(K_r,\bar{Q}))^{m_r}\nonumber\\
&\hspace{-2in}=&  \hspace{-1in}\prod_{r=1}^L 
(v_1^{GB}(K_r,\bar{Q}))^{n m_r}
\;\; .\eeqar{vn2ell}
This  approximate mean recoil factorization and remarkable power scaling
of coherent state semi-classical \br wave harmonics leads
to an apparent ``perfect fluid collective flow'' interpretation.

Higher order cumulant harmonic correlations were
 proposed~\cite{vnflow,Bzdak:2013rya,ATLASvn2k, vncumurefs}
to help remove ``non-flow'' sources of correlations such as 
momentum conservation, back to back dijet, and Bose statistics effects
and isolate true collective bulk fluid flow azimuthal asymmetries.
The
$2\ell$-particle cumulant suppresses ``non-flow'' contribution by eliminating the correlations which act between fewer than $2\ell$ particles
(see. e.g., fig.9 of \cite{ATLASvn2k}).
The first few cumulants for $2\ell=2,4,6$ (notation from
from Ref.~\cite{Bzdak:2013rya,ATLASvn2k})
are
\begin{widetext}
\beqar
(v_n\{2\})^2&\equiv& \left\langle 
e^{i n(\phi_1- \phi_2)}\right\rangle \equiv \langle|v_2|^2\rangle \nonumber\\
(v_n\{4\})^4&\equiv& \left\langle -
e^{i n(\phi_1+\phi_2-\phi_3-\phi_4)} \right\rangle + 2 \left\langle 
e^{i n(\phi_1-\phi_3)} \right\rangle \left\langle 
e^{i n(\phi_2-\phi_4)} \right\rangle = 2\langle |v_2|^2\rangle^2
-\langle |v_n|^4\rangle \nonumber\\
(v_n\{6\})^6 &\equiv& \left\langle 
e^{i n(\phi_1+\phi_2+\phi_3-\phi_4-\phi_5-\phi_6)} \right\rangle - 
9 \left\langle |v_2|^2\right\rangle \left\langle |v_n|^4\right \rangle +12\left\langle |v_2|^2\right\rangle^3)/4
\; \; . \eeqar{bzdak}
\end{widetext}
The observed~\cite{ATLASvn2k} near equality of $v_n\{2 \ell\}$ for $\ell=2,3,4$
in $Pb+Pb$ at LHC has been interpreted as strong evidence for 
perfect fluid flow.  The similarity of ``elliptic flow'' $v_2\{4\}(p_T)$
in p+Pb and Pb+Pb observed by ATLAS\cite{Aad:2012gla}
and also for ``triangular flow'' $v_3\{4\}(p_T)$ by CMS\cite{CMS:2012qk}
has been interpreted as evidence for
perfect fluidity also on sub-nucleon scales in p+Pb.

However, we show that color \br exhibits the same scaling of cumulants
in the mean recoil approximation from a single antenna.
In the case that all $2\ell$ momenta are identical,  
\beq
\bar{v}_n\{2\ell\} 
\equiv (v_m^{M=1}\{2\ell\}(k,\cdots,k;\bar{Q} ))^{n/m}
\eeq{pureplow}
which in the previous notation implies, for example, that
\beqar
\langle |v_n|^4\rangle&=&\left\langle |v_2|^2\right\rangle^2 \\
\langle |v_6|^6 \rangle &=& \langle |v_2|^2\rangle \langle |v_n|^4\rangle=
\left\langle |v_2|^2\right\rangle^3
\eeqar{GBrules}
and similarly for all cumulants.
Therefore color \br obeys the {\em the same} azimuthal
harmonic cumulant independence on the 
number of gluons $2\ell$ used to determine the harmonic moments
as does perfect fluid local equilibrium. 
However in our case, 
the ``flow'' effect comes purely from zero temperature
pure coherent state (semi-classical) wave interference effects 
produced from the geometric arrangement of CSA arrays .  

For the case of multiple $M>1$ independent target cluster
CSA arrays the cumulant harmonic moments depend in a more complex
way on the particular geometric and recoil correlations defining the 
CSA. Special analytic cases for $v_2\{2\}$ considered corresponding 
to idealized $Z_n$ and
Gaussian CSA arrays are illustrated  in the following two sections.

\section{Example of Special {\protect $Z_n$} CSA Array \Br}
\begin{figure}[!ht]
\subfigure[]{\includegraphics[width=3.in]{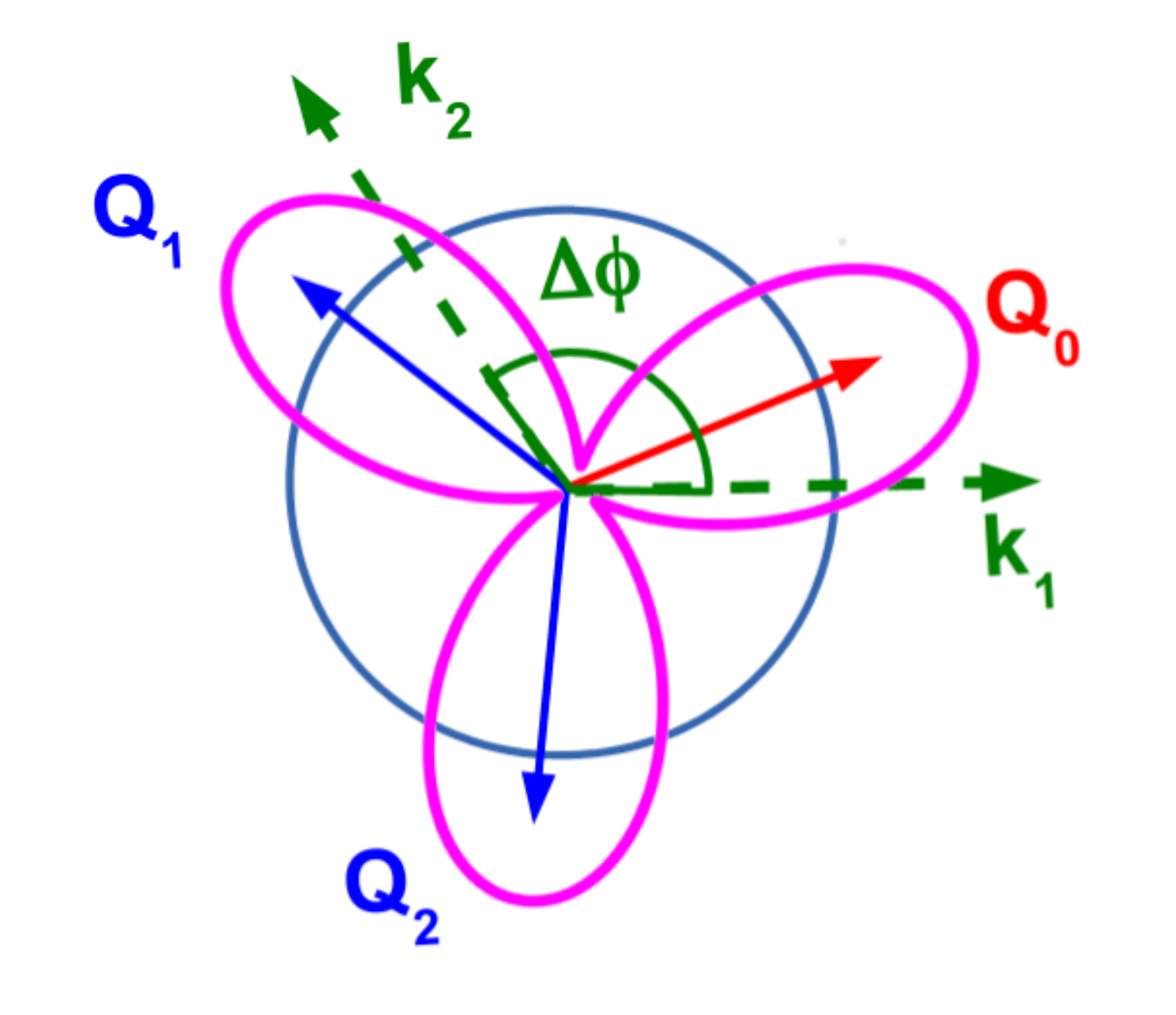}}\\
\subfigure[]{\includegraphics[width=1.5in]{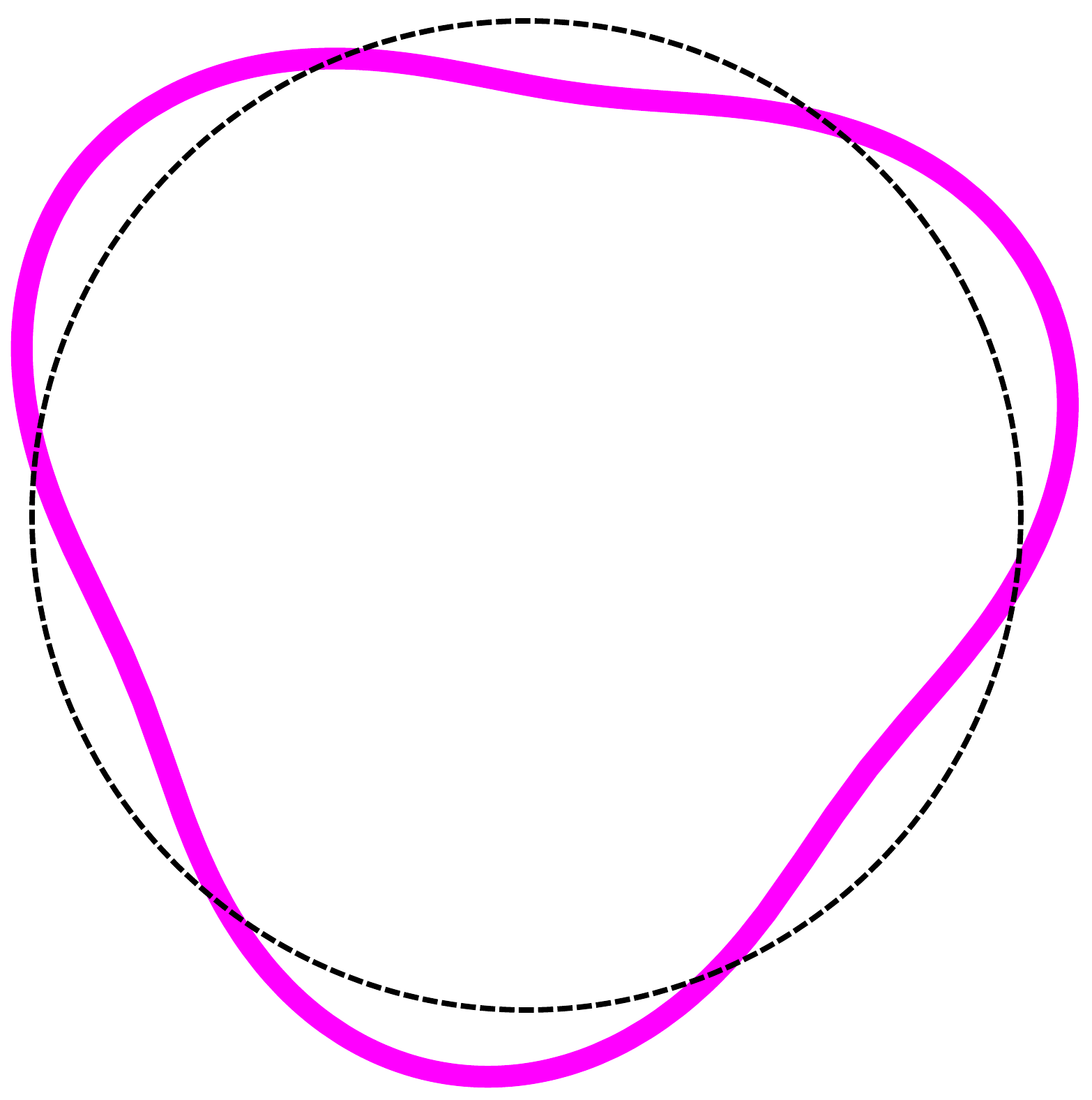}}
\caption{(Color online) Example illustrating apparent perfect ``triangular 
flow'' using $Z_3$ 
Color Scintillation Antenna arrays with $M=2$
target clusters recoiling off projectile
with ${\bf Q}_0= -\sum_{a=1}^M {\bf Q}_a$. All $Q_a$ are assumed
to have same magnitude but 
spaced by $2\pi/3$ in azimuthal angle. $Z_n$ CSA radiate  ``perfect'' 
factorized two particle $n^{th}$
harmonics $v_{n}\{2\}(k_1,k_2)=\delta_{n,3}\;v_3^{GB}(k_1,Q_0)v_3^{GB}(k_2,Q_0)$.
Part (a) shows extreme case $v_3=0.45$ while  
(b) is for more realistic $v_3=0.07$. The isotropic soft non perturbative 
background is assumed to be subtracted out.   
}  
\label{fig-3}
\end{figure}

\begin{figure}[!ht]
\subfigure[]{\includegraphics[width=3.in]{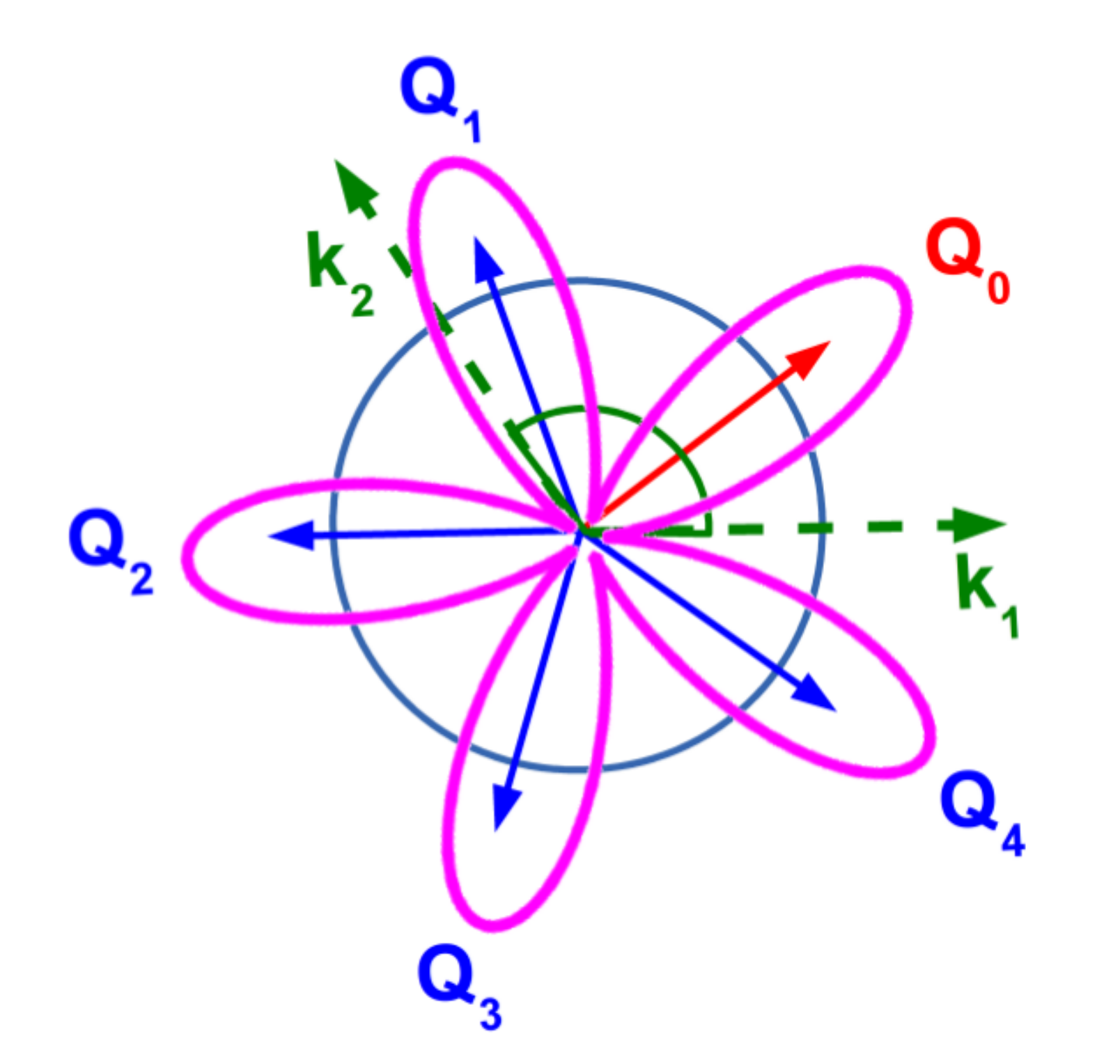}}\\ 
\subfigure[]{\includegraphics[width=1.5in]{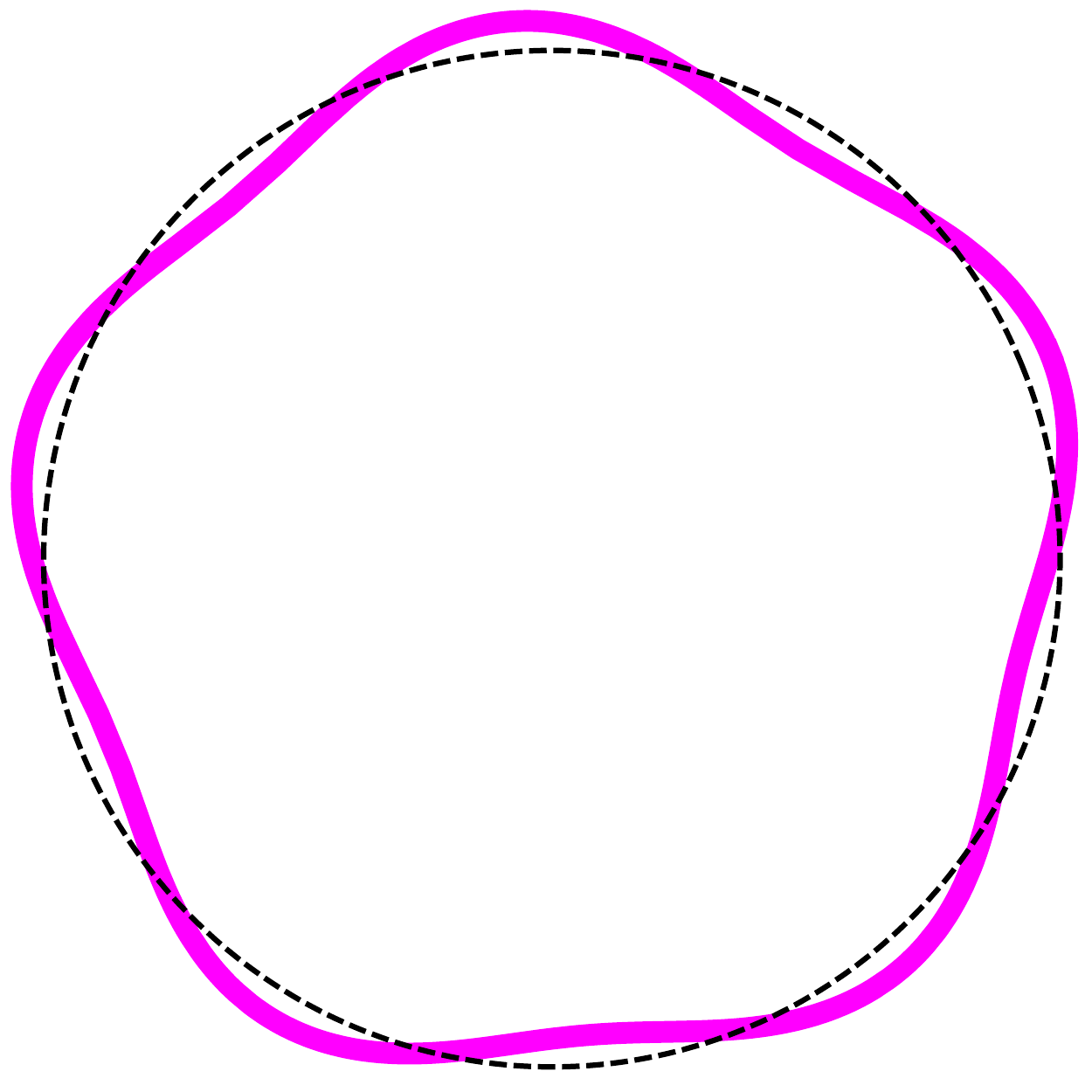}}
\caption{(Color online) As in Fig.~\ref{fig-3} but for a $Z_5$ CSA
producing apparent perfect pentatonic flow 
$v_{n}\{2\}(k_1,k_2)=\delta_{n,5}\;v_5^{GB}(k_1,Q_0)v_5^{GB}(k_2,Q_0)$. 
Part (a) show extreme $v_5=0.45$ while part (b) shows more realistic
$v_5 = 0.03$.
}  
\label{fig-4}
\end{figure}

As see in Appendix B, from Eqs.~(\ref{2fnMN}) it is clear that a particularly simple
special cases of color antenna arrays where $M=n-1$ target beam jet 
clusters all have similar number of recoiling target partons 
$m_a=N/M=N/(n-1)$ and for which transfer all $n=M+1$ projectile and target
 beam jets recoil   with similar momentum transfers,  $Q_a^2=N/M \mu^2$, 
but with specially spaced azimuthal angles, $\{\psi_a\}= 2\pi a/n$.

These particular  color antenna arrays that we will refer to as $Z_n$ Color Scintillation Arrays (CSA) have a special
 discrete  azimuthal rotation symmetry corresponding to the finite
 group of $n$ roots of unity, 
\beq Z_n=\left \{z_{a,n}=e^{i 2\pi  a/n}|a=0,\cdots,n-1; \sum_{a=0}^{n-1}z_{a,n} =0\right \}
\; \; . \eeq{un}
For these $Z_n$ CSA  geometries 
of projectile and target color dipole antennas
the double sum over $a$ and $b$ is trivial because
\beq
\cos(n(\psi_a-\psi_b))= \cos(2\pi(a-b))=1
\; \; , \eeq{rou1}
and thus all $(M+1)^2=n^2$ terms are identical. Note that Eq.(\ref{rou1})
is invariant to global $SO(2)$ simultaneous rotations of all antennas.

What is remarkable about $Z_{M+1}$ symmetric CSA is that due to the
orthogonality properties of the $z_{an}$ phases, 
\beqar
\sum_{a=1}^{n-1} z_{a,n}^k &=& n \delta_{k,n} \\
\sum_{a=1}^{n-1} (z_{a,n})^k  (z^*_{a,n})^{k'} &=& n \delta_{k,k'}
\; \; ,\eeqar{znsum}
 all harmonics except
 $n=M+1$ vanish! The $Z_n$ CSA thus scintillate with ``perfect'' $n$-harmonic
azimuthal correlations.  For $Z_n$ CSA 
the two particle relative Fourier moments $v_n\{2\}$   simply factor into a product
of single particle moments $v^{GB}_n(k_i,Q_0,0)$ because
the $n$  complex ${Q}_a=Q_0 z_{a,n}$  
form a regular polygon with equal radii as illustrated for an
 $n=5$ ``star fish'' antenna array in 
Fig.~(\ref{fig-3}) that generate a perfect $\cos(5(\phi_1-\phi_2))$ two particle
azimuthal correlation.

For roots of unity CSA color antenna geometries all $M+1$ 
antennas receive the same $Q_a^2=Q_0^2 = N/(n-1) \mu^2$ momentum transfer
and produce the same
 single particle $v_{M+1}^{GB}(k,Q_0,0)$ harmonics. 
Since the two particle harmonics vanish except for $n=M+1$,
\beqar
v_n^{M,N}\{2\}(k_1,k_2) &\stackrel{Z_n}{\longrightarrow}& \delta_{n,M+1} v_{M+1}^{GB}(k_1,Q_0) v_{M+1}^{GB}(k_2,Q_0)   \nonumber \\
\frac{v_n^{M,N}\{2\}(k_1,k_2)}{ v_{M+1}^{GB}(k_2,Q_0)} &\stackrel{Z_n}{\longrightarrow}& \delta_{n,M+1} \;(v_1^{GB}(k_1,Q_0))^{M+1} \; , 
\eeqar{2vnfactored}
and for $n=M+1$ , $v_{M+1}^{M,N}\{2\}(k_1,k_2)$ is reduced to simply the product
of single GB CSA moments at $k_1$ and $k_2$.

Examples of $Z_n$ radiation patterns for $n=3,5$ 
for  extreme high $v_n=0.45$ in parts (a) and more realistic $v_3=0.7$ and $v_5=0.03$ from Fig.~(1) are shown in Figs.~(\ref{fig-3},\ref{fig-4}).

\section{Gaussian Color Scintillation Antenna arrays}
Another simple limit is when the recoil azimuthal angles $\psi_a$  are in random $[0,2\pi]$ and the ${\bf Q}_a$ are distributed 
with a Gaussian of same width squared $\langle Q_a^2\rangle=Q_T^2=(N/M) \mu^2$  
for $ a\in[1,\cdots,M]$.  
In this antenna array, the projectile ${\bf Q}_0$ is also  Gaussian distributed 
with zero mean but with an enhanced second moment, 
\beq
\langle Q_0^2\rangle= M Q_T^2 =N\mu^2 
\;\; . \eeq{q02ean}
Unlike for perfect $n^{th}$ harmonic antenna arrays with Eq.~(\ref{rou1}),
in the random Gaussian distributed case 
\beq
\cos(n(\psi_a-\psi_b))= \delta_{a,b} 
\; \; , \eeq{rou2}
and so only the $a=b$ diagonal terms contribute.
All $a\ge 1$ target terms are identical and only the projectile contribution
is enhanced due to $\langle Q^2_0\rangle/Q_T^2=M$ random walk exchanges from each cluster.
In this case, Eq.(\ref{2fnMN}) reduces to 
\begin{widetext}
\beqar 
 f^{N,M}_n({k}_1,{k}_2) & \stackrel{Gauss}{\rightarrow}& 
\int d^2{\bf Q} \left\{\frac{\exp[-{\bf Q}^2/(2N\mu^2)]}{2\pi N \mu^2} 
+ M \frac{ \exp[-{\bf Q}^2/(2(N/M)\mu^2)]} {2\pi (N/M) \mu^2} \right\}
\left\{ B_{1Q}B_{2Q}
\; 
f_{0,1,Q}\;f_{0,2,Q} \right.\nonumber\\
&\;& \hspace{3.50in}\left. \; \times v_n^{GB}(k_1,Q)\; v_n^{GB}(k_2,Q) \right\}
\;\; ,\\
 f^{N,M}_n({k},{k}) & \stackrel{Gauss}{\rightarrow}& 
\int d^2{\bf Q} \left\{\frac{\exp[-{Q}^2/(2N\mu^2)]}{2\pi N \mu^2}
+ M \frac{\exp[-{Q}^2/(2(N/M)\mu^2)]}{2\pi (N/M) \mu^2}\right\} 
\left\{B_{kQ}f_{0,k,Q} v_n^{GB}(k,Q)\right\}^2
\;\; .\eeqar{sca2}
\end{widetext}
We have suppressed target and projectile kinematic rapidity factors.

To get a feeling for the magnitude of the two particle azimuthal moments
we can approximate $Q$ in the  integrand outside the Gaussian weights by its
rms $\Delta Q=\sqrt{\langle Q^2 \rangle}$ and perform the normalized integral over the Gaussians to estimate
\begin{widetext}
\beqar 
\sqrt{{f^{N,M}_n({k},{k})}} & \approx&  \left(\frac{C_R \alpha_s\mu^2}{\pi^2k^2} \right)
 \left\{
 \frac{1}{(N+1)\mu^2}\frac{(v_1^{GB}(k,\sqrt{N}\mu))^{n}}{((k^2+ (N+1)
 \mu^2)^2 - 4Nk^2\mu^2)^{1/2}}\;\;   \right. \nonumber \\
&\;& \hspace{0.9in} \left. +  \frac{M}{(N/M+1)\mu^2}
\frac{(v_1^{GB}(k,\sqrt{N/M}\mu))^{n}}{((k^2+ (N/M+1) \mu^2)^2-
4(N/M)k^2\mu^2)^{1/2}}\right \}
\;\; .\eeqar{sca2a}
\end{widetext}
The rapidity dependence corresponding to 
the BGK\cite{Brodsky:1977de}
 triangular rapidity enhancement $N ({Y_P -\eta})/({Y_P-Y_T})$
of the single inclusive multiplicity toward the target fragmentation region
is suppressed above to simplify the result. In addition we emphasize that
the mostly non-perturbative low $k$ background is ignored in our simplified consideration here.
Full account for that background will require implementation of the above
non-isotropic soft \br in an event generator such as HIJING.

A qualitative BGK\cite{Brodsky:1977de} rapidity dependence for target cluster number $M(\eta)$ 
that ignores  the $c/k$ resolution scale considerations
discussed in Eq.~(\ref{percol}) can be estimated 
by identifying $N=\chi=\int dz \rho_A(z,{\bf b})$ with the opacity
as a function of $b$ and taking
\beq
M_{BGK}(\eta)  \sim \chi ({Y_P -\eta})/({Y_P-Y_T})(1-e^{Y_T-\eta})^{n_f} 
\;\;. \eeq{rapbgk}

The main feature expected from 
 BGK\cite{Brodsky:1977de} rapidity dependence of the target
cluster \br  is that the $v_n\{ 2\ell \}$ cumulant harmonics
depend on the $\eta_i$  in a characteristic way that
reflects underlying triangular BGK $p+A$ inclusive boost non-invariant
inclusive rapidity distribution. There is also a rather 
nontrivial combined $(\eta_i,k_i)$ variation of 
the moments due to the peaking of
the $v_n^{GB}(k,Q)$ near $k^2=Q^2$ 
and the fact that target cluster rapidity dependent recoil 
$k^2= (N/M(\eta))\mu^2$ varies with $\eta$ while  the fully coherent 
Cronin enhanced $k^2\sim N\mu^2$ projectile peak increases with the
opacity $N\sim L/\lambda=\chi$. 
 Detailed numerical studies of combined $(\eta,k)$ dependence of 
azimuthal color \br  will be explore elsewhere.

\section{HIJING Monte Carlo Color Scintillating Beam Jets}

To get a realistic estimate for the magnitudes and systematics
of pQCD VGB induced harmonics  in realistic $p+p, p+A, A+A$ collisions, we have to embed the
anisotropic recoil bremsstrahlung gluons into phenomenological
Lund strings Schwinger hadronization scheme that has been
tuned to reproduce low $p_T$ {\em $\phi$ averaged} inclusive 
hadronic observables in  $e^+ + e^-$, $e+p$, $p+p$, $p+A$, as well as $A+A$. HIJING Monte Carlo event generator 
is one such model based on the LUND~\cite{Andersson:1986gw} string model and
PYTHIA and JETSET~\cite{Sjostrand:1993yb} Monte Carlo models.

Simple local parton-hadron duality prescription as used in CGC cannot be expected to predict quantitative
hadron mass dependent moderate $p_T<2$ GeV anisotropy moments over three decades of $\sqrt{s}$.
 The advantage of Monte Carlo event generators built on multi-decade
phenomenological analysis is that they summarize  the world data by taking into account the particle data book, quantum number, and energy momentum 
conservation and numerous Standard Model dynamical details.
Of course, they do not proport to cover all possible phenomena.

A key feature  missing in HIJING and most other event generators for $A+B$ collisions
so far  are basic pQCD azimuthal anisotropies at the moderate
$p_T<2$ GeV scale that are so clearly predicted by GB and generalized
VGB bremsstrahlung models.   What has been included in most event generators 
are strong jet anisotropies due to collinear factorized
pQCD mini and hard jet production above a scale $p_T> p_0\sim 2$ GeV. 
As currently implemented, HIJING only take into 
softer scale $k<p_0$ gluons as random transverse string ``wiggles'' 
using ARIADNE~\cite{Ariadne} Lund model scheme, but it neglects entirely the basic pQCD recoil correlations so 
explicitly seen in the GB and VGB bremsstrahlung expressions. 
An open question is the magnitude of the anisotropies that will arise 
when the ARIANDE part of the JETSET code 
is replaced by VGB anisotropic bremsstrahlung as derived here.
We intend to address this numerically intensive work  as a future application of the formulas derived in this paper.

\section{Conclusions}

In summary,  we applied the GLV reaction operator approach to the
Vitev-Gunion-Bertsch (VGB) boundary conditions to compute the
all-order in nuclear opacity non-abelian gluon \br of event-by-event
fluctuating beam jets in nuclear collisions. 
We obtained analytic expressions for 
the azimuthal Fourier cumulant moments $v_n\{2\ell\}$ as a function
of the gluons kinematics ${\bf k}_i,\eta_i\}$   
in terms of single gluon beam jet GB \br harmonics.
These moments obey remarkably simple  power law scaling 
similar to the ones observed recently 
in high energy p+A reactions at RHIC and LHC 
as a function of the target participant
clusters geometry.  Multiple clusters of projectile
and target beam jets form Color Scintillation Antenna (CSA) arrays 
that radiate gluons with characteristic
boost non-invariant trapezoidal rapidity distributions in asymmetric
$B+A$ nuclear collisions.  The intrinsically azimuthally anisotropic 
and long-range in $\eta$ nature of the non-abelian \br 
leads to  $v_n$ moments that are similar to results from perfect fluid 
hydrodynamic models,
but due entirely to non-abelian wave interference phenomena
sourced by the fluctuating CSA.  
We presented examples of simple solvable models of 
target dipole clusters and showed that  our analytic non-flow 
\br solutions for $v_n\{2\ell\}$  
are similar to recent numerical saturation model predictions 
but differ by predicting
a simple power-law hierarchy of both even and odd $v_n\{2\ell\}$ without
invoking details of $k_T$ factorization, though 
CGC saturation evolution
is expected to be important for future quatitative comparisons to data.
The basic CSA mechanism can be tested via its
predicted systematic boost non-invariant
 $\eta$ rapidity dependence in $B+A$

Non-abelian beam jet CSA \br  investigated in this paper may  
provide a partial analytic solution to the  Beam Energy
Scan (BES) discovery of the near $\sqrt{s}$ 
independence of the azimuthal moments
down to very low CM energy of $\sim$ 10 AGeV,  
where large $x$ valence quark beam jets
dominate inelastic dynamics. Recoil \br from multiple independent
CSA clusters also provides a natural pQCD qualitative explanation
for the unexpected  similarity of $v_n$ in $p(D)+A$ and
non-central $A+A$ at same $dN/d\eta$ multiplicity as observed at RHIC and LHC.

This PQCD based model show that the uniqueness of perfect fluid descriptions of
$p+A$ and $B+A$ data cannot be taken for granted. However, a great deal 
of work remains to sort out the fraction of the observed $v_n\{2 \ell\}$ azimuthal harmonic systematics that can be properly 
ascribed to final state bulk collective flow versus QCD coherent state
color scintillating wave interference phenomena.

\section{Appendix: The Linked Cluster Theorem for coherent VGB Gluon Bremsstrahlung}

To derive the link cluster theorem for the coherent limit of VGA
 we introduce the
shorthand notation for the integrations over momentum transfers
\beq
\prod_{j=1}^n\int d(w_j-\delta_j)\equiv\int \prod_{j=1}^n d^2 {\bf q}_j    
 \left( \frac{1}{\sigma_{el}} 
\frac{d \sigma_{el}}{d^2 {\bf q}_j}   
-  \delta^2 ({\bf q}_j) \right)
\; ,
\eeq{wdel}
which have the convenient properties $\int dw_j=\int d\delta_j=1$ and  
$\int d(w_j-\delta_j)=0$, that is particularly  useful to be able to discard any terms in the integrand that
does not depend simultaneously on all $n$  ${\bf q}_j$ momenta
at fixed opacity order $n$. Using this shorthand and ${\bf C}_{jn}$ notation from Eqs.\ref{defsC},
we rewrite the right hand side of Eq.~(\ref{VGBf}) as
\begin{widetext}
\beqar
VGB &=& 
\frac{C_R \alpha_s}{\pi^2}\sum_{n=1}^\infty \frac{\chi^n}{n!}
\left[\prod_{j=1}^n \int d( w_j -\delta_j)\right]  
\nonumber \\ 
&\;& \hspace*{1cm} \times  \; 
\left( {\bf C}_{2n}-{\bf C}_{1n}\right)
\cdot 
\left[ \left( {\bf C}_{2n}-{\bf C}_{1n}\right)
+2 \left({\bf C}_{3n}-{\bf C}_{2n}\right)+\cdots +
2 \left({\bf C}_{(n+1)n}-{\bf C}_{nn}) \right]\right. 
\nonumber \\
&=& 
\frac{C_R \alpha_s}{\pi^2} \sum_{n=1}^\infty \frac{\chi^n}{n!}
\left[\prod_{j=1}^n \int d( w_j -\delta_j) \right]   
\left( {\bf C}_{2n}-{\bf C}_{1n}\right)
\cdot 
\left[ \left( {\bf C}_{2n}-{\bf C}_{1n} \right)
+2 \left({\bf H}-{\bf C}_{2n})\right)
\right]
\nonumber \\
&=& 
\frac{C_R \alpha_s}{\pi^2} \sum_{n=1}^\infty \frac{\chi^n}{n!}
\left[\prod_{j=1}^n \int d( w_j -\delta_j)\right] 
\left[ -({\bf H}-{\bf C}_{2n})+({\bf H}-{\bf C}_{1n})\right]
\cdot 
\left[  ({\bf H}-{\bf C}_{2n})+({\bf H}-{\bf C}_{1n}) 
\right] \nonumber\\
&=& 
\frac{C_R \alpha_s}{\pi^2} \sum_{n=1}^\infty \frac{\chi^n}{n!}
\left[  \prod_{j=1}^n \int d( w_j -\delta_j) 
\right] 
\left\{ |{\bf H}-{\bf C}_{1n}|^2 
- |{\bf H}-{\bf C}_{2n})|^2
 \right\}
\nonumber\\
&=& 
\frac{C_R \alpha_s}{\pi^2} \sum_{n=1}^\infty \frac{\chi^n}{n!}
\left[  \prod_{j=1}^n \int d( w_j -\delta_j) \right] |{\bf H}-{\bf C}_{1n}|^2 
\nonumber\\
&=& 
\sum_{n=1}^\infty \frac{\chi^n}{n!}
\left[  \prod_{j=1}^n \int d( w_j -\delta_j) 
\right] 
\left(\int d^2{\bf Q}\;\delta^2({\bf Q} - ({\bf q}_1+\cdots +{\bf q}_n))\right)
\left\{\frac{C_R \alpha_s}{\pi^2} \frac{{\bf Q}^2}{k^2({\bf k}-{\bf Q})^2}\right\}
\; .
\eeqar{VGBcoh}
\end{widetext}
Here, we used the notation ${\bf H}\equiv{\bf C}_{(n+1),n}\equiv {\bf k}/k^2$
to denote the ``hard'' vacuum radiation amplitude that
shows up at zeroth order in opacity in the case final state induced radiation
in GLV~\cite{Gyulassy:2000er}. Note that in this notation
 convention  ${\bf B}^n_{(n+1), n)} \equiv {\bf H}-{\bf C}_{nn}$.  

Note that $\int d(w_j-\delta_j)=0$, and therefore $j=1$ integral of $-|{\bf H}-{\bf C}_{2n}|^2$ automatically vanishes.
Note further that  the $|{\bf H}-{\bf C}_{1n}|^2$ integrand 
depends only on ${\bf k}$ and the {\em Total} accumulated ${\bf Q}=\sum_{i=1}^n {\bf q}_i$ momentum transfer. Thus, the integrand is symmetric under arbitrary permutations if the indices. This is the key to obtain the linked cluster
rearrangement because out of the 
the  $2^n$ combinations of the $w_j$ and minus delta functions $-\delta_i$, 
all combinations with the same number $m$ of $\int dw$ and $n-m$ of $\int d \delta$
integrations give the same contribution.  At fixed opacity order $n$
the $2^n$ combinations of integrals reduce to sum over only $n$  integrals 
of the form $n!/(m!(n-m)!) \int dw_1\cdots dw_m (-1)^{n-m}|B^m_{1m}|^2$. 
Therefore, 
\begin{widetext}
\beqar
\frac{dN^{VGB}_{coh}}{d\eta d^2 {\bf k} }
&=& 
\sum_{n=1}^\infty \frac{\chi^n}{n!} \sum_{m=1}^n \frac{(-1)^{n-m}\;n!}{m!(n-m)!}
\int d^2{\bf Q} \left[  \int dw_1\cdots dw_m \delta^2({\bf Q} - ({\bf q}_1+\cdots +{\bf q}_m))\right]   
\left\{\frac{C_R \alpha_s}{\pi^2} \frac{Q^2}{k^2({\bf k}-{\bf Q})^2}\right\}
\; . \quad
\eeqar{VGBlct}
\end{widetext}
Changing summation variables from, $\infty> n\ge 1$ and $n\le m\ge 1$ 
to $\infty> \ell=n-m\ge 0$ and $\infty> m \ge 1$,
the double sum $\sum_{\ell=0}^\infty\sum_{m=1}^\infty$ factorizes, and
 the sum over $\ell$  produces a factor $\exp[-\chi]$ corresponding to the probability of no scattering.
Therefore, Eq.(\ref{VGBlct}) leads to link cluster theorem 
Eq.(\ref{linkclusterth}) for the multiple collision VGB generalization of 
Gunion-Bertsch gluon \br.

\section{Appendix B:  Two gluon \br azimuthal harmonics $v_n\{2\}$ }

For the two gluon case azimuthal harmonic 
correlations can be directly derived in another way 
 by integrating 
over both $\phi_1=\Phi+\Delta\phi/2$ and $\phi_2=\Phi-\Delta\phi/2$ keeping the 
relative azimuthal angle 
$\Delta\phi= \phi_1-\phi_2$ fixed and weighing the integrand by $\cos(n\Delta \phi)$ from
\begin{widetext}
\beqar 
 f^{M}_n\{2\}({k}_1,{k}_2) &\equiv &
\int_{-\pi}^\pi \frac{d\Phi}{2\pi} \,   \int_{-\pi}^\pi \frac{d\Delta\phi}{2\pi}  \, \cos(n \Delta\phi)  
dN^{M}_2(k_1,\Phi+\Delta\phi/2,k_2,\Phi-\Delta\phi/2) \nonumber\\
&\hspace{-1in}=& \hspace{-0.6in}
\sum_{a,b=0}^M B_{1a}B_{2b}\int_{-\pi}^\pi  \frac{d\Delta\phi}{2\pi}  \,  \cos(n \Delta\phi) 
\int_{-\pi}^\pi       \frac{d\Phi}{2\pi}  
\frac{1}{A_{1a}-\cos(\Phi+\psi_a+\Delta\phi/2)}
\frac{1}{A_{2b}-\cos(\Phi+\psi_b-\Delta\phi/2)} \label{2vn1}\\
&\hspace{-1.0in}=&  \hspace{-0.6in}
\sum_{a,b=0}^M  B_{1a}B_{2b} \int_{-\pi}^\pi        \frac{d\Phi'}{2\pi}         \,  \frac{1}{A_{1a}-\cos(\Phi')}
\int_{-\pi}^\pi    \frac{d\Delta\phi}{2\pi} \, 
\frac{ \cos(n \Delta\phi)}{A_{2b}-\cos((\Phi'+\psi_b-\psi_a) -\Delta\phi)} 
\label{2vn2}\\
&\hspace{-1.0in}=&   \hspace{-0.6in}
\sum_{a,b=0}^M B_{1a}B_{2b}\; f_{n,2,b} \int_{-\pi}^\pi    \frac{d\Phi'}{2\pi}     \, 
\frac{\cos(n(\Phi'+\psi_b-\psi_a))}{A_{1a}-\cos(\Phi')}  
=\sum_{a,b=0}^M B_{1a}B_{2b}\;f_{n,2,b} \;f_{n,1,a} \;\cos(n(\psi_b-\psi_a)) \label{2vn5}\\
&\hspace{-1.0in}=&   \hspace{-0.6in} \sum_{a,b=0}^MB_{1a}B_{2b}\;   
f_{0,1,a}\;f_{0,2,b} \; (v_1^{GB}(k_1,Q_a)v_1^{GB}(k_2,Q_b))^n \; \cos(n(\psi_b-\psi_a)) \label{2fnMN}
\;\; ,\eeqar{sca2b}
\end{widetext}
where we defined
$\Phi'=\Phi+\psi_a+\Delta\phi/2$ and used periodicity of the integrand to shift the 
$\Phi'$ range back to $[-\pi,\pi]$ in  Eq.~(\ref{2vn2}), then performed 
the $\Delta\phi$ integral with the help of Eq.~(\ref{vn1}).
We used here the shorthand notation
\beq
f_{n,1,a}=\int_{-\pi}^\pi  \frac{d\Phi}{2\pi}  \ \frac{\cos(n \Phi)}{A_{1a}-\cos(\Phi)}
=(v_1^{GB}(k_1,Q_a))^n f_{0,1,a}, \eeq{fn1a}
\beq
f_{n,1,a}=\frac{\left(A_{k_1,Q_a}-\sqrt{A^2_{k_1,Q_a}-1}\right)^n }{\sqrt{A^2_{k_1,Q_a}-1})} 
\eeq{fn1a2}
\beq
 \lim_{\mu\rightarrow 0}\; f_{n,1,a} =
\left(\frac{k_1}{Q_a}\right)^n\frac{\theta(Q_a-k_1)}{Q_a^2-k_1^2} 
Q_a^2  \; .
\eeq{fn1a3}

\section{Acknowledgments} 
MG  is  grateful to W. Busza,  J. Harris, J. Jia, A. Poszkanzer, H.J. Ritter, N. Xu
for  discussion related to RHIC and LHC flow experiments, and to  
A. Dumitru, T. Lappi,  L. McLerran, J. Noronha, H. Stoecker, G. Torieri, 
and R. Venugopalan for critical discussions related to hydrodynamic and
QCD field theory models of A+A correlations.  MG acknowledges support from the 
US-DOE Nuclear Science Grant No.\ DE-FG02-93ER40764, partial  sabbatical support
from LBNL under DOE  No.\ DE-AC02-05CH11231, the Yukawa Institute for Theoretical Physics, Kyoto 
University, the YITP-T-13-05 on "New  Frontiers in QCD" workshop support, and sabbatical support from the
MTA Wigner RCP, Budapest, where this work was finalized. PL and TB
acknowledge support from Hungarian OTKA grants  K81161, K104260, NK106119, and
NIH TET\_12\_CN-1-2012-0016.   IV was supported in part by the US Department of Energy, 
Office of Science.


\end{document}